\documentclass{aastex}
\usepackage{emulateapj5,epsfig}

\makeatletter

\newenvironment{inlinefigure}{%
\def\@captype{figure}%
\noindent\begin{minipage}{0.999\linewidth}\begin{center}}
{\end{center}\end{minipage}\smallskip}
\makeatother

\begin{document}

\def\wdf{white dwarf}
\def\etal{et al.} 
\def\rd{Di\thinspace Stefano}

\title{Luminous Supersoft X-Ray Sources in External Galaxies}

\author{R.~Di\,Stefano} 
\affil{Harvard-Smithsonian Center for Astrophysics, 60
Garden Street, Cambridge, MA 02138}
\affil{Department of Physics and Astronomy, Tufts
University, Medford, MA 02155}
\author{A. K. H. Kong} 
\affil{Harvard-Smithsonian Center for Astrophysics, 60
Garden Street, Cambridge, MA 02138}

\begin{abstract}

We use a set of conservative criteria to identify 
luminous supersoft X-ray sources
(SSSs) in external galaxies. 
We test this approach on blackbody 
models, and then apply it to {\it Chandra}
data from $4$ galaxies: 
an elliptical (NGC 4967), $2$ face-on spirals
(M101 and M83), and an interacting galaxy (M51). 
 We find SSSs in every galaxy, 
and estimate that the
total SSS population
of sources with $L > 10^{37}$ erg s$^{-1}$
in each galaxy is at least several hundred and could be significantly larger.  
In addition, we discover two intriguing features of
galactic populations of SSSs. First, there are significant
subpopulations of high-luminosity sources; the
 bolometric luminosities
can exceed 
$10^{39}$ erg s$^{-1}.$ 
Second, 
in   
the spiral galaxies M101, M83 and M51, 
SSSs appear to
be associated with the spiral arms. This may indicate that
some SSSs are young systems, possibly younger than $10^8$ years.

\end{abstract}


\section{Introduction}

Luminous supersoft X-ray sources (SSSs)
are characterized by values of  
$k\, T$ on the order of tens of eV, and 
luminosities greater than $\sim 10^{35}$ erg s$^{-1}$.
The corresponding 
effective radii are comparable to those of white dwarfs (WDs).
Indeed, some hot WDs and pre-WDs have been observed as SSSs
(see Greiner 2000),
including several recent novae, symbiotic systems,
 and a planetary nebula (PN). 
More than half
of all SSSs with optical IDs, however, do not seem to be examples
of systems of known types. It is the mysterious nature of these
other systems that has excited much of the interest in SSSs.
A promising model 
is one in which matter from a Roche-lobe-filling
companion 
accretes onto the WD at rates so high ($\sim 10^{-7} M_\odot$ yr$^{-1}$)
that it can undergo quasi-steady 
nuclear burning (van den Heuvel et al. 1992;
Rappaport, \rd\ \& Smith 1992; Hachisu, Kato \& Nomoto 1996,
\rd\ \& Nelson 1996). 
Because matter that is burned can be retained by the WD, 
some SSS binaries may be
progenitors of Type Ia
supernovae.  

There is indirect evidence 
that nuclear-burning WDs (NBWDs) 
with a Roche-lobe-filling donor  
exist in nature. The majority of the SSS binaries
whose nature has not yet been determined are considered to be
candidates for the ``close-binary SSS" (CBSS) model, in which
mass transfer is driven on the thermal time scale of a donor
whose Roche lobe shrinks as mass transfer proceeds.
 Nevertheless, studies
of the $7$ Magellanic Cloud and $2$ Milky Way
candidates for this or other NBWD models 
have not yet found direct evidence for the presence of a WD
or of a donor.
We do know, however,
that an X-ray pulsar may present itself an SSS. RX J0059.2-7138
has a non-pulsed luminous ($\sim 6.7\times 10^{38}$ ergs s$^{-1}$),
 soft ($\sim 30$ eV) component (Hughes 1994). 
Observers outside of the beam of the hard pulsed radiation 
would
classify the source as an SSS. 
This example serves to illustrate the danger of assuming that
all SSSs 
whose natures we do not yet understand  
are NBWDs. We also note that the stripped cores of 
giant stars that have been tidally disrupted by massive
black holes are expected to
 appear as SSSs for times
ranging from $10^3$ to $10^6$ years (\rd\ et al. 2001). 
Several such stripped cores could be present within 
$\sim 1$ kpc of the nuclei of galaxies harboring high-mass black holes.  
  
Because gas obscures $> 99\%$ of the Milky Way's SSSs, 
studies of the size and characteristics of galactic
populations of SSSs can be carried out only
in external galaxies.
Here we apply a single set of conservative criteria, applicable to
both {\it Chandra} and XMM data, 
to identify SSSs in $4$ galaxies, allowing direct comparisons
of their SSS populations.

\subsection{A Phenomenological Definition of SSSs}  

Most SSS spectra are similar to blackbody spectra, although the
inclusion of WD atmosphere effects
has produced some improved fits. 
Early definitions of the class
assumed $k\, T < 50$ eV, but this upper  
limit 
is unrealistic,  since even the flagship
SSS CAL 87, has $63$ eV $< k\, T < 84$ eV (Greiner 2000). 
In total, $4$
of the $9$ CBSS candidates have estimated upper limits on $k\, T$
larger than $50$ eV, and $2$ also have lower limits
above $50$ eV, as does the Galactic recurrent nova U Sco,
a system which has almost certainly experienced an episode
of nuclear burning. 
There are also theoretical arguments for an upper limit above $50$ eV.
For example, WDs with masses close to the Chandrasekhar limit are expected to
have $k\, T$ near $100$ eV. To derive an extreme upper limit,
we note that a $1.4\, M_\odot$ WD
with Eddington-luminosity nuclear burning on its
surface would have $k\, T \sim 150$ eV.
Because the effective radius is likely to be
larger than the radius of the WD,
expected temperatures are lower, unless some of these  
systems can circumvent the Eddington limit.   
  
There is an independent reason to place the division between
SSSs and and other X-ray sources at
$k\, T \sim 175$ eV (\rd\ \& Kong 2003).
This is 
that  {\it Chandra's} detector response (which is
similar to that of {\it XMM} in the soft energy regime), 
is such that sources below this temperature
do not contribute to radiation in the ``hard" waveband
above $2$ keV, while sources with $k\, T \le\ 175$ eV do
exhibit a hard X-ray component (see \rd\ \& Kong 2003).   

If there are sources with temperatures greater than $\sim 125$ eV
but less than $175$ eV, these are interesting in any case.
They could represent  (1) an extension 
of the class of SSSs, (2) low-T supernova remnants (SNRs);
these can be distinguished from X-ray binaries
through variability studies, or (3) a new class of intermediate temperature
X-ray binaries.

The following phenomenological definition 
selects from the pool of all X-ray sources detected
in an external galaxy, the majority of those sources
which are SSSs,
regardless of their physical nature.   
An X-ray source is an SSS if either of the following conditions are met.
{\bf (1)} The spectrum is well-fit by a 
blackbody model with $k\, T < 175$ eV,  
or {\bf (2)} 
Whatever the best fit model, 
fewer than $10\%$ of the energy
is carried by photons with energy greater than $1.5$ keV.
The reason for this condition is that, even if the  
spectrum starts out as a pure blackbody spectrum,
interactions of the radiation with matter near the source,
or wind interactions (expected for many SSSs that are quasi-steady NBWDs)
can add a small harder component.
\footnote{The choice of $10\%$
is rather arbitrary; future studies of larger numbers of
SSSs may suggest that this criterion be altered. For now
we allow as much as $10\%$ of the energy to come in at energies
above $1.5$ keV, because this is roughly
consistent with efficient reprocessing or with, e.g.,
 the
presence of a companion emitting a strong shocked wind.
We do not allow higher values, simply because we want to
reserve the SSS classification for  sources whose spectrum is clearly
dominated by the soft component.}

Most sources in external galaxies provide too few photons for a
meaningful spectral fit. 
In M101, e.g., $51$ of $118$ sources provide fewer than $30$
photons, $40$ sources provide fewer than $25$ photons,
and $27$ provide fewer than $20$ photons.  
It is therefore important to design a uniform procedure
to identify those SSSs (the majority) providing fewer than $\sim 200$ photons.  
The selection procedure introduced and applied in this paper 
preferentially selects sources with $k\, T < 100$ eV.   

\section{Selection Criteria} 

\subsection{The HR Conditions}

In this paper we will employ strict hardness ratio 
(HR) criteria to identify SSSs. We use 
$3$ energy bins to define hardness ratios: 
{\bf S:} 0.1-1.1 keV, {\bf M:} 1.1-2 keV, {\bf H:} 2-7 keV. 
We consider $2$
hardness ratios, HR1 and HR2, demanding that   
\begin{equation} 
{{HR}1} ={{M-S}\over{M+S}} < -0.8
\end{equation} 
and 
\begin{equation} 
HR2 ={{H-S}\over{H+S}} < -0.8. 
\end{equation} 
These conditions imply that  
$S > 9\, M$,  
and $S > 9\, H$.  

We also introduce $2$ additional hardness ratios that include
uncertainties. 
\begin{equation}
HR1_{\Delta} =
{{(M+\Delta M)-(S+\Delta S)}\over{(M+\Delta M)+(S+\Delta S)}} < -0.8, 
\end{equation}
and 
\begin{equation}
HR2_{\Delta}=
{{(H+\Delta H)-(S+\Delta S)}\over{(H+\Delta H)+(S+\Delta S)}} < -0.8, 
\end{equation} 
where 
$\Delta S$, $\Delta M$, and $\Delta H$ are the one-$\sigma$ uncertainties in
$S$, $M$, and $H,$ respectively. 
Conditions (3) and (4) 
strengthen the criteria, implying that
$(S + \Delta S) > 9\, (M + \Delta M)$,  
and $(S + \Delta S) > 9\, (H + \Delta H)$.  
Sources satisfying these $4$ conditions are denoted ``SSS-HRs".

For sources that do not provide many photons, the
precise definitions of $\Delta\, S,$ $\Delta\, M,$ and $\Delta\, H$
play a crucial role in determining which X-ray sources are tagged as 
SSS-HRs. 
There are $2$ sources of uncertainty: background, and
counting statistics. 
For each of the $3$ bins, we use $\sigma^2=\sigma'^2(N_S)+ N_B,$
where $N_B$ is the number of counts extracted from a
background region of area equal to that of the source region,
$N_S$ is the background-subtracted number of photons 
extracted from the source region, and $\sigma'(N_S)$ is the one-$\sigma$
uncertainty in $N_S$.
Since, for {\it Chandra} observations, the background is generally
negligible, the sizes of  $\Delta\, S,$ $\Delta\, M,$ and $\Delta\, H$
can depend primarily on the value of $\sigma'(N_S)$ for each bin.   
For small values of
$N_S,$ $\sigma'(N_S)=1+\sqrt{N_S+0.75}$; for large values of
$N_S,$ we have  $\sigma'(N_S)=\sqrt{N_S}.$  
We have chosen to represent $\sigma'(N_S)$ by a function
that (1) is exactly
equal to the appropriate small-$N_S$ values
for $N_S = 0, 1,$
(2) approaches $\sqrt{N_S}$, for large
$N_S,$ and (3) interpolates between the  
two at intermediate values. The interpolation formula gives a value of $\sigma$ near 2 for $N_S <5$ and closely approximates $\sqrt{N_S}$ for $N_S >5$

Given these prescriptions, the designation SSS-HR requires that 
 the $S$ band receive
more than $13.1$   photons if no photons arrive in either the $M$ 
or $H$ bands,
while $S$ must receive more than $24$ photons if there is even a
single photon in either $M$ or $H.$   

\subsection{Selected Sources}

In cases in which $N_H$ is below $\sim 2 \times 10^{21}$ cm$^{-2}$,
and in which the count rate is high enough that photon statistics
do not significantly alter the values of the hardness ratios, 
the HR conditions will select most SSSs with spectral
profiles like those of the nearby SSSs (Greiner 2000).
For example, a $100$ eV source behind a column of 
$1.6 \times 10^{20}$ cm$^{-2}$ will easily satisfy the
HR conditions if it  provides enough counts.
Several circumstances can, however, make this or other true SSSs
fail to fulfill the
HR criteria.
For example, a low count rate, combined with the vagaries 
of photon statistics will 
sometimes cause a large enough deficit of photons below $1.1$ keV,
and a also a large enough excess of 
photons
with energies above $1.1$ keV, to 
cause a true SSS to fail the HR conditions.
Thus, as the tests described below illustrate, the HR conditions
will select only the strongest candidates SSSs.

\subsection{Tests}

We performed a set of simulations,
applying the HR criteria to each of a large number of
blackbody sources used to seed galaxies similar to 
those we will study in \S 3.
Each seeded source was characterized by $3$ physical parameters:  
$k\, T$  ($25$ eV $\le\ k\, T \le 2$ keV), 
$L$ ($6.9 \times 10^{35}\le L_{0.3-7\, keV} 
\le 1.4 \times 10^{40}$ erg s$^{-1}$),
and $N_H$ 
($4 \times 10^{20}$ erg s$^{-1} \le N_H < 2.5 \times 10^{22}$ cm$^{-2}$).
We used the PIMMS software (AO3 release) to
compute the counts that would be detected in a $50$ ksec ACIS-S
observation if 
each physical source 
were located in a galaxy $5$ Mpc, $10$ Mpc, and $15$ Mpc away. 
Using PIMMS in ACIS-S mode is equivalent to assuming that 
each source is located near the center of 
the backside illuminated S3 chip. 

For each physical source and for each energy bin
 we used $10$ values of a random variable $r$
to choose the number of photons detected:
$S-2\, \Delta\, S < S_{detected} < S + 2\, \Delta\, S,$ etc.
Choosing uniformly over the $\pm 2-\sigma$ interval
slightly overestimates
the effects of the uncertainties. 

Each galaxy was seeded with 12920 X-ray sources, of which 6080
were SSSs.
We applied the HR conditions only to
sources that provided more than $14$ photons.
Figure 1 is a graphical representation of the results
for a galaxy located at $10$ Mpc,
seeded
with sources with a maximum luminosity,
$L_{max},$ of $1.4 \times 10^{40}$ erg s$^{-1}$.
The region for which the selection criteria are
most relevant is the one located between the horizontal lines.
(The upper horizontal line corresponds to $200$ counts. Sources
providing more counts can be fit by spectral models, so a selection
procedure is less crucial.) 
Examination of this region
reveals that the HR conditions work very well for 
$55$ eV $\le\ k\, T \le\ 100$ eV. For low to moderate absorption
they select a fair fraction of sources at $25$ eV, and $40$ eV, and
$125$ eV, and only a few unabsorbed sources with $k\, T = 150$ eV.

As Figure 1 indicates, the HR conditions
can successfully identify high-count ($> 200$ count) SSSs with   
 little contamination
from non-SSSs.
Since, however, the conditions are not strictly needed for these
high-count cases, we present in Table 1 test results only
for those cases in which the selection procedure
can work ($> 14$ counts) and in which it is needed ($< 200$ counts).
The numbers of SSSs providing between $14$ and $200$ counts 
were $1225$ for the galaxy at $5$ Mpc, 
$1106$ for the galaxy at $10$ Mpc,
and $1021$ for the galaxy at $15$ Mpc. 
In each galaxy, approximately $1/2$ of all the seeded SSSs 
with counts in the range $14-200,$   
were selected as SSSs by the HR conditions.  
(See Table 1.) Furthermore, among the sources identified as SSS-HRs,
the level of contamination was minimal. With more than $500$ true
SSSs identified as SSS-HRs, there were $4$, $7$, and $3$,
outliers mistakenly identified as SSSs
 in the galaxies at $5,$ $10$, and $15$ Mpc,
respectively. All of these  outliers were
located behind small gas columns ($4 \times 10^{20}$ cm$^{-2}$),
had small count rates, and satisfied the 
HR conditions only because of the effects of
limited photon statistics.  
Most of the outliers had 
$k\, T = 175$ eV,
while $3$ of the $14$ had $k\, T = 200$ eV;
thus, even the outliers were remarkably soft X-ray sources.

Examination of both Figure 1 and Table 1 reveal that
the HR conditions do have a weakness: they fail to
identify a significant fraction of true SSSs. For large values of
$L_{max}$ ($1.4 \times 10^{40}$ erg s$^{-1}$), roughly half of
all SSSs sources yielding between $14$ and $200$  counts were  
not identified as SSS-HRs. Although the most heavily absorbed sources
were most likely to be missed (particularly if $k\, T$ was 
 near
$100$ eV), sources with less absorption and low-temperature
sources were also missed by the HR criteria. The simulation
carried out with a smaller value of $L_{max}$ 
($1.0 \times 10^{38}$ erg s$^{-1}$),
demonstrates that the fraction of SSSs not selected is larger for
low-luminosity sources, particularly in more distant galaxies,
even though only sources providing 
more than $14$ counts were considered. 

\begin{table*}
\caption{Tests of the HR Conditions}
{\centering
\footnotesize
\begin{tabular}{lccccc}
\hline
\hline
 & \multicolumn{2}{c}{$L_{max}=1.4\times10^{40}$ erg s$^{-1}$} &&
\multicolumn{2}{c}{$L_{max}=10^{38}$ erg s$^{-1}$} \\
\cline{2-3} \cline{5-6}\\
Distance & $N^{SSS}_{HR}$$^{(a)}$ & $N^{SSS}_{not\ HR}$$^{(b)}$ &&
$N^{SSS}_{HR}$$^{(a)}$ & $N^{SSS}_{not\ HR}$$^{(b)}$
\\
\\
\hline
D=5 Mpc  & 598 (4)$^{(c)}$ & 627 && 310 (4) & 337\\
D=10 Mpc & 554 (7) & 552 && 147 (3) & 212\\
D=15 Mpc & 513 (3) & 508 && 52 (3) &133\\
\hline
\end{tabular}
\par
\medskip
\begin{minipage}{0.8\linewidth}
\footnotesize

For each galaxy distance
(rows: $5$ Mpc, $10$ Mpc, $15$ Mpc),
two sets of simulations were carried out.
In one set of simulations,
$L_{max} = 1.4\times10^{40}$ erg s$^{-1}$,
where $L_{max}$ is the maximum luminosity of any
seeded source.
In the second set of simulations, $L_{max}= 10^{38}$ erg s$^{-1}$.

$^{(a)}$ $N^{SSS}_{HR}$ is the number of SSS-HRs with
between $14$ and $200$ counts.\\
$^{(b)}$ $N^{SSS}_{not\ HR}$ is the number of SSSs with
between $14$ and $200$ counts {\it not} identified as SSS-HRs.\\
$^{(c)}$ The numbers in parentheses are the numbers of
sources that are {\it not} SSSs, but which are identified
as SSS-HRs. All such mis-identified sources have
$k\, T \le\ 200$ eV and lie behind small
($4 \times 10^{20}$ cm$^{-2}$) hydrogen columns.
\end{minipage}
\par
}
\end{table*}

\subsection{Real Chandra Data}

{\it Chandra's} ACIS
detectors have suffered degradation at low ($<$ 1 keV) energies.
This degradation is time dependent and, as it is presently understood,
it was not included in the
PIMMS AO3 simulations. In the next section, we apply the
selection algorithm to real {\it Chandra} ACIS-S data taken
at different times since 1999. For some of the observations,
 the AO3 release of
PIMMS provides an adequate guide 
to the flux as a function of temperature
and $N_H$.
The progressive low-energy
degradation has the effect, however, of making
our selection  procedures more and more strict when applied to
galaxy data taken later and later into
{\it Chandra's} mission. 
For example, when the 
effects of the low-energy degradation cause there to be only $2/3$  
as many counts in the $S$ bin
as there would have been in the simulated data,
while $M$ and $H$ are relatively unaffected, imposing
the requirement that
$S > 9\, M$, will be the same as if we had imposed
$S > 13.5\, M$ in the simulations. That is, the low-energy degradation
makes it more difficult for a true SSS to be identified as such.
There is no direct cure
for this problem, because we cannot assume that
a lack of low-energy photons is due to a lack of sensitivity.
Fortunately, even with the degradation, we are able to identify
a significant population of SSSs in each galaxy.

\begin{inlinefigure}
\psfig{file=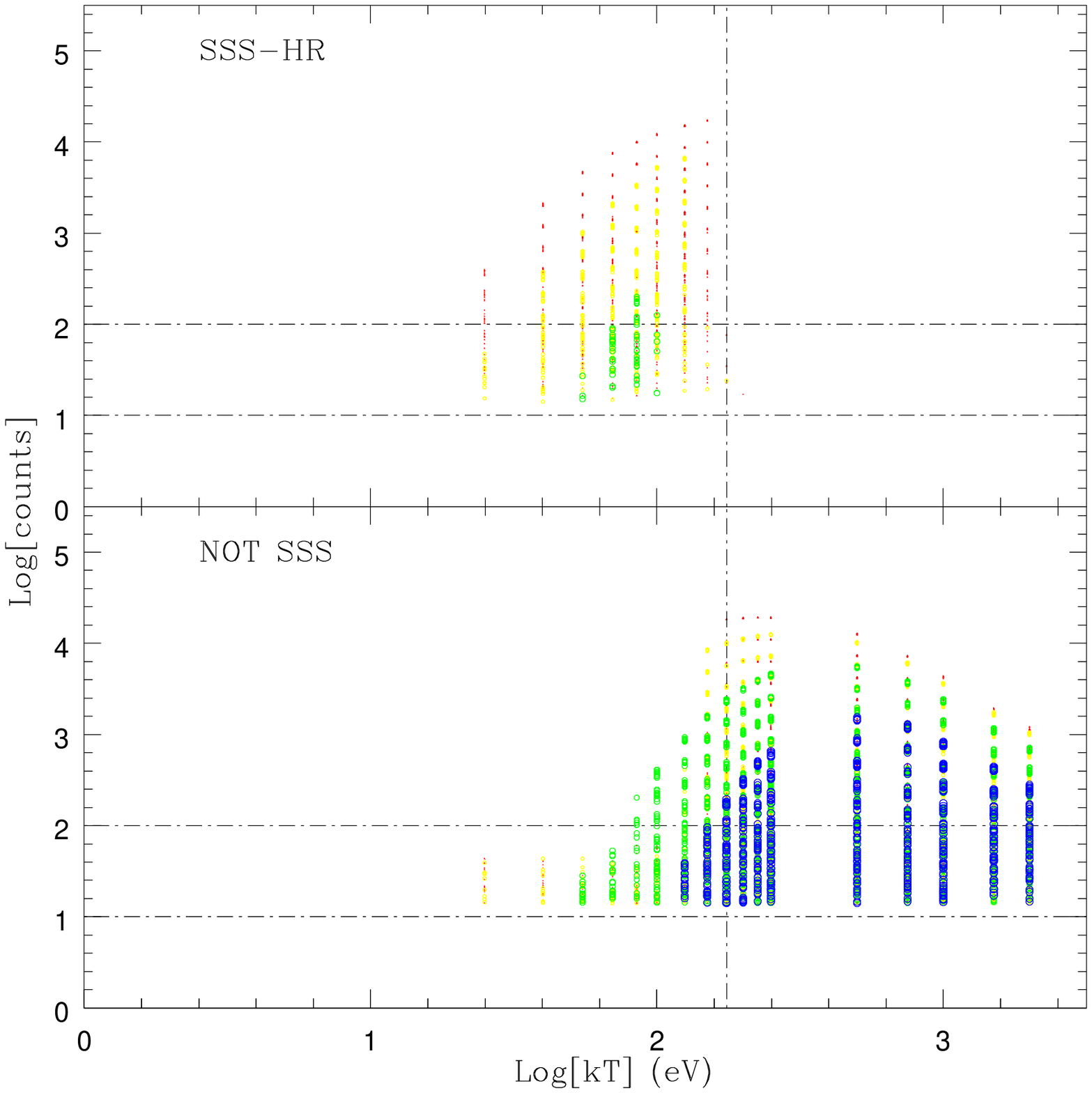,height=3in}
\caption{Test of the HR criteria as applied to
thermal models. Along the vertical
axis is the log of the number of counts that PIMMS 
(AO3 release) predicts would be detected
for sources located in a galaxy $10$ Mpc away; statistical uncertainties
have been included.
 Source luminosities ranged from
$6.9 \times 10^{35}$ ergs s$^{-1}$
to $1.4\times10^{40}$ erg s$^{-1}$.
Four values of $N_H$ were considered:
$4.0 \times 10^{20}$ cm$^{-2}$ (red points),
$1.6 \times 10^{21}$ (yellow; open circles) cm$^{-2}$,
$6.4 \times 10^{21}$ cm$^{-2}$ (green; larger open circles),
$2.5 \times 10^{22}$ cm$^{-2}$(blue; largest open circles).
The logarithm of the temperature is
plotted along the horizontal axis. Upper panel: points are shown only for
those sources that would be identified as SSS-HRs.
Lower panel: sources
not selected as SSSs by the HR conditions.
The first vertical line of    sources on the left have $k\, T = 25$ eV;
$k\, T$  increases by $15$ eV through $100$ eV, then in $25$ eV intervals
through to the vertical line at $175$ eV, denoting the
temperature cut-off for SSSs.
}
\end{inlinefigure}

Many distant galaxies are large enough that studies of the
X-ray source  population must analyze
the data taken in all of the chips, not just the
backside illuminated chip. For example, a galaxy of diameter
$30$ kpc, located $10$ Mpc away, has dimensions of about
$20'$, and is therefore spread out over at least $3$ chips. Even
if the galaxy is smaller or more distant, halo
studies would include chips in addition to S3.
Unfortunately, however, the soft X-ray sensitivity
(which varies somewhat even within S3), is dramatically
smaller outside of the backside-illuminated chip.
An AO3 ACIS-I observation of a $50$ eV ($80$ eV) source located
behind a column of $4 \times 10^{20}$ cm$^{-2}$, e.g., will 
collect only $\sim 14\%$ ($20\%$) as many photons   
as if the sources were located on-axis in S3.     
This inevitably implies that some low-flux SSSs will fail
to produce a significant detection in the $S$ band. 
Such sources are lost to any reasonable selection algorithm.
 
\subsection{{\sl XMM} Observations}

The HR criteria were designed to optimize
the selection of SSSs as simulated by PIMMS for {\it Chandra}
ACIS-S observations. As presently formulated, they will have
almost the same effect on data from {\sl XMM},
which offers similar energy
coverage but with larger effective area.
We compared
the sensitivity of {\sl
XMM}/PN (with a thin filter) and {\it Chandra} ACIS-S.
For SSSs 
 we found that, in 
the S, M, and H bands,
{\sl XMM} collects $\sim 3$, $\sim 2$, and
$\sim 3$ times as many photons as
{\it Chandra}, respectively.
This suggests that the primary difference between
{\sl XMM}/PN and {\it Chandra} ACIS-S is that the former has somewhat
smaller relative sensitivity to photons in the $M$ band. 
{\sl XMM} observations of distant galaxies do suffer from
significantly more background than {\it Chandra}
observations of the same galaxies. 
Nevertheless,
because of its larger effective area, and also because of the
degradation in {\it Chandra's} low-energy sensitivity, 
{\sl XMM} may do better in searching for SSSs
in regions where source confusion and diffuse gas emission
are not problems.

\subsection{Foreground and background objects}

The field of each galaxy contains X-ray sources unrelated to the
galaxy, such as 
background AGN, and 
X-ray active foreground stars.
Given the exposure times of the galaxies we study in \S 3,
we may expect $10-15$ foreground/background objects
in M51, M83, and NGC 4697 (see, e.g., Soria \& Wu 2002; 
Sarazin, Irwin, \& Bregman 2001), and close to $30$ in the field
of M101 (Pence et al.\ 2001). For the purposes of this paper,
we must estimate the fraction of the foreground and background
objects likely to be identified by the HR conditions as SSSs.
To make this estimate, 
we have investigated 
data from the ChaMP ({\it Chandra}
Multiwavelength Project \footnote{http://hea-www.harvard.edu/CHAMP})
archives (P. Green, private communication). In 5 ACIS-S observations
with durations $\sim 10-20$ ks, 
no sources satisfied the HR conditions.
Therefore, even though there are soft AGN, and even though many
X-ray active stars emit soft X-rays, such sources
appear unlikely to significantly contaminate the pool of SSS-HR
sources.

\section{Application to $4$ Galaxies}

We applied the HR conditions to the X-ray sources in $4$ galaxies:
$3$ spirals and an elliptical. 
Table 2 summarizes key properties of the galaxies,
and of the {\it Chandra} observations.
M101 lies along a
direction with little Galactic absorption and our view of
the galaxy is almost face on; it is therefore an ideal place
to search for SSSs.   M83 is almost face-on and has experienced recent
star bursts. M51 is more inclined, and is interacting with another
galaxy.  
For M101 and M51 we generated our own source lists with the CIAO tool
WAVDETECT. For M83 we used a source list provided by R. Soria,
and for NGC 4697 we used the source list of 
Sarazin, Irwin, \& Bregman (2001). 
If the lower limit of the  energy range that had originally been
used by WAVDETECT 
was higher than $0.1$ keV, we carried out a visual inspection to
search for any sources with reliable detections but with few 
photons detected above $0.3$ keV.   
Details, and 
complete source lists can be found in \rd\ \& Kong (2003b). 
We found $33$ SSS-HR sources in the $4$ galaxies: $3\%$ of the 
sources detected in NGC 4697 are SSS-HRs, while the comparable fraction 
for M101 is $15\%$ (see Table 3).

\begin{table*}
\caption{Summary of sample galaxies}
{\centering
\footnotesize
\begin{tabular}{lccccccccl}
\hline
\hline
Name& Type& Distance & $D_{25}$ & Inclination & $N_H$ & {\it
Chandra} & $M_{BH}$ &$N_{SSS}/N_{total}$$^a$ \\
    &     &  (Mpc)   &  (arcmin)&                &
($10^{20}$cm$^{-2}$) & exposure (ks) &$(\times 10^{8}
M_{\odot})$ &\\
\hline
M101 & Sc & 5.4, 6.7$^1$ & 23.8 & $0^{\circ}$ & 1.2 & 94.4, 9.6$^b$
 &$< 0.01$ $^2$& 16/118\\
M51  & Sc$^c$ & 7.7, 8.4$^3$ & 13.6 & $64^{\circ}$ &
1.6 & 14.9, 26.8$^b$ & 0.1\,$^4$ & 3/71; 2/92\\
M83 &  Sc & 4.57$^5$, 4.7 & 11.5 & $24^{\circ}$ & 3.8 &
49.5& 0.1\,$^6$ & 10/128\\
NGC 4697 & E & 11.7$^7$, 15.9$^8$, 23.3 & 7.1  &
$44^{\circ}$$^d$ & 2.1 & 39.3 & 1.2$^9$ & 3/91\\
\hline
\end{tabular}
\par
\medskip
\begin{minipage}{0.8\linewidth}
\footnotesize

NOTE. --- All data are from Nearby Galaxies Catalogue (Tully 1988)
unless specified. M101 was observed on 2000 March 26 and 2000 October 29$^b$; M51 was observed on 2000 June 20 and 2001 June 23$^b$; M83 was observed on 2000 April 29; NGC 4697 was observed on 2000 January 15.\\

$^a$ Ratio of number of SSSs to total number of X-ray sources.\\ 
$^b$ The second observation.\\
$^c$ Seyfert 2 galaxy.\\
$^d$ For elliptical galaxies
the inclination is given by $3^{\circ}$ +
acos\,$\left(\sqrt{((d/D)^2 - 0.2^2)/(1 - 0.2^2)}\,\right)$, where d/D
is the axial ratio of minor to major diameter (Tully
1988). 
This inclination angle is generally unrelated to the value of $N_H$.\\

REFERENCES. --- 1: Freedman et al. 2001; 2: Moody et al. 1995; 3: Feldmeier
et al. 1997; 4: Hagiwara et al. 2001; 5:
Karachentsev et al. 2002; 6: Thatte et al. 2000; 
7: Tonry et al. 2001; 8: Faber et al. 1989; 9: Ho 2002

\end{minipage}
\par
}
\end{table*}

\begin{table*}
\caption{SSS-HR source list}
{\centering
\footnotesize
\begin{tabular}{lcccccccccc}
\hline
\hline
\multicolumn{1}{c}{Object}& R.A.& Dec. & \multicolumn{2}{c}{Soft} & \multicolumn{2}{c}{Medium} & \multicolumn{2}{c}{Hard} & HR1$^a$ & HR2$^b$\\
\cline{4-5} \cline{6-7} \cline{8-9}\\
 & (h:m:s)& $(^{\circ}:\arcmin:\arcsec)$ &Counts & S/N & Counts & S/N & Counts & S/N&&\\
\hline
M101-15(10) & 14:02:59.4&+54:20:42&   31.3 &   5.3 &   1.3 &   0.9 &   0.0 &   0.0 &  -0.9(-0.9)  & -1.0(-0.9)\\
M101-18(13*) & 14:03:01.1&+54:23:41&  169.6 &  12.9 &   0.0 &   0.0 &   0.1 &   0.1 &  -1.0(-1.0)  & -1.0(-1.0)\\
M101-21(16*) & 14:03:02.5&+54:24:16&   74.2 &   8.4 &   0.0 &   0.0 &   0.0 &   0.0 &  -1.0(-1.0)  & -1.0(-1.0)\\
M101-34(27) & 14:03:07.9&+54:21:23&   35.7 &   5.8 &   1.8 &   1.2 &   0.0 &   0.0 &  -0.9(-0.9)  & -1.0(-1.0)\\
M101-37(30*) & 14:03:08.5&+54:20:57&    29.3 &   5.1 &   1.0 &   1.0 &   0.0 &   0.0 &  -0.9(-0.9)  & -1.0(-0.9)\\
M101-38(31) & 14:03:08.6&+54:23:36&   20.7 &   4.1 &   0.0 &   0.0 &   0.0 &   0.0 &  -1.0(-1.0)  & -1.0(-1.0)\\
M101-43(36) & 14:03:10.6&+54:21:26&   18.2 &   3.8 &   0.0 &   0.0 &   0.0 &   0.0 &  -1.0(-1.0)  & -1.0(-0.9)\\
M101-50(43) & 14:03:13.2&+54:21:57&   31.3 &   5.3 &   0.0 &   0.0 &   0.4 &   0.3 &  -1.0(-1.0)  & -1.0(-0.9)\\
M101-51(45*) & 14:03:13.6&+54:20:09&  219.1 &  14.6 &   0.3 &   0.2 &   0.0 &   0.0 &  -1.0(-1.0)  & -1.0(-1.0)\\
M101-55(48) & 14:03:13.9&+54:18:11&   34.3 &   5.6 &   1.4 &   1.0 &   0.0 &   0.0 &  -0.9(-0.9)  & -1.0(-0.9)\\
M101-78(73) & 14:03:22.6&+54:20:38&   34.3 &   5.3 &   1.5 &   1.0 &   1.0 &   0.5 &  -0.9(-0.9)  & -0.9(-0.9)\\
M101-80(75) & 14:03:24.0&+54:23:37&   68.7 &   8.0 &   0.8 &   0.5 &   0.0 &   0.0 &  -1.0(-0.9)  & -1.0(-1.0)\\
M101-97(92) & 14:03:29.8&+54:20:58&   71.9 &   7.7 &   0.0 &   0.0 &   0.0 &   0.0 &  -1.0(-1.0)  & -1.0(-0.9)\\
M101-101(96*) & 14:03:31.9&+54:23:23&   30.9 &   5.0 &   0.0 &   0.0 &   0.0 &   0.0 &  -1.0(-1.0)  & -1.0(-1.0)\\
M101-102(98) & 14:03:32.3&+54:21:03& 8512.0 &  92.2 & 658.4 &  25.6 &  68.0 &   8.0 &  -0.9(-0.9)  & -1.0(-1.0)\\
M101-102 (2nd obs) & & &290.2 & 17 & 4.7 & 2.1 & 0.5 & 0.5 &-1.0 (-1.0) & -1.0 (-1.0)\\
M101-104(99*) & 14:03:33.3&+54:18:00&  227.8 &  14.5 &   0.3 &   0.1 &   1.1 &   0.3 &  -1.0(-1.0)  & -1.0(-1.0)\\
\\
M83-20 & 13:36:53.9&-29:48:48&   43.0 &   5.6 &   2.6 &   1.2 &   2.6 &   1.0 &  -0.9(-0.8)  & -0.9(-0.8)\\
M83-42 & 13:36:59.1&-29:53:36 &   30.5 &   5.3 &   0.0 &   0.0 &   0.0 &   0.0 &  -1.0(-1.0)  & -1.0(-1.0)\\
M83-50 &13:37:00.4&-29:50:54 &   105.1 &  10.0 &   0.8 &   0.7 &   0.0 &   0.0 &  -1.0(-1.0)  & -1.0(-1.0)\\
M83-54 & 13:37:01.1&-29:54:49 &  132.6 &  11.4 &   1.8 &   1.3 &   0.7 &   0.7 &  -1.0(-1.0)  & -1.0(-1.0)\\
M83-79 & 13:37:06.1&-29:52:32 &  108.6 &  10.3 &   1.9 &   1.3 &   0.0 &   0.0 &  -1.0(-0.9)  & -1.0(-1.0)\\
M83-81 & 13:37:06.5&-29:54:16&   15.7 &   3.7 &   0.0 &   0.0 &   0.0 &   0.0 &  -1.0(-1.0)  & -1.0(-1.0)\\
M83-88 & 13:37:07.4&-29:51:33 &   17.5 &   3.9 &   0.0 &   0.0 &   0.0 &   0.0 &  -1.0(-1.0)  & -1.0(-0.9)\\
M83-98 &13:37:12.8&-29:50:12&     35.9 &   5.8 &   2.1 &   1.4 &   1.6 &   1.1 &  -0.9(-0.8)  & -0.9(-0.9)\\
M83-111 & 13:36:59.5&-29:52:03&   78.2 &   8.4 &   5.9 &   2.2 &   2.0 &   1.3 &  -0.9(-0.8)  & -0.9(-0.9)\\
M83-128 & 13:36:57.7 & -29:53:53 & 29.6 & 5.3 & 1.6 & 1.1 & 0.0
&0.0 &-1.0(-1.0) & -0.9(-0.8)\\
\\
M51-12 & 13:29:43.3&+47:11:34&  272.7 &  16.5 &   5.2 &   2.1 &   1.0 &   1.0 &  -1.0(-0.9)  & -1.0(-1.0)\\
M51-12 (2nd obs)& &                    &  366.0 &19.1 & 11.4 & 3.4 & 0.0 & 0.0 & -0.9(-0.9) & -1.0(-1.0)\\ 
M51-42 & 13:29:55.4&+47:11:43&   14.7 &   3.6 &   0.0 &   0.0 &   0.0 &   0.0 &  -1.0(-0.9)  & -1.0(-1.0)\\
M51-58 & 13:30:02.3&+47:12:38&   20.4 &   4.4 &   0.0 &   0.0 &   0.0 &   0.0 &  -1.0(-1.0)  & -1.0(-1.0)\\
M51-51 (2nd obs) & 13:29:55.3&+47:11:26 & 14.6 & 3.4 & 0.0 & 0.0 & 0.0 & 0.0 & -1.0(-1.0) &-1.0(-1.0)\\
\\
NGC4697-16 & 12:48:37.1&-05:47:58 &   63.0 &   7.9 &   0.0 &   0.0 &   0.0 &   0.0 &  -1.0(-1.0)  & -1.0(-1.0)\\
NGC4697-19 & 12:48:34.5&-05:47:49&   73.7 &   8.5 &   0.9 &   0.9 &   0.0 &   0.0 &  -1.0(-1.0)  & -1.0(-1.0)\\
NGC4697-52 & 12:48:41.2&-05:48:19&   58.6 &   7.5 &   0.0 &   0.0 &   0.7 &   0.7 &  -1.0(-1.0)  & -1.0(-0.9)\\
\hline
\end{tabular}
\par
\medskip
\begin{minipage}{0.8\linewidth}
\footnotesize
NOTE.--- The object ID in parentheses is from Pence et al. (2001);
SSSs selected by Pence et al. (2001) are noted by ``*''.

$^a$ $HR1=(M-S)/(M+S)$; value in parentheses is HR1$_{\Delta}$.
$^b$ $HR2=(H-S)/(H+S)$; value in parentheses is HR2$_{\Delta}$.

\end{minipage}
\par
}
\end{table*}

\subsection{Spectra} 

Eight sources provided enough photons to allow spectral fits.
(Two of these sources permitted spectral fits in each of $2$
separate observations.) 
In NGC 4697, the most distant galaxy, we considered
a composite of all $3$ SSS-HRs, which happened to have similar
spectral profiles  
in bins of $0.2$ keV from $0.1$ to $0.7$ keV.

Before presenting the results, we note that there is likely to
be more systematic uncertainty in fits  of SSS spectra than in the
fits of spectra associated with any other class of sources.
Calibration issues have not been resolved at the softest energies,
introducing obvious anomalies into the spectra  of bright nearby
SSSs. Most SSSs in distant galaxies
 provide too few photons for these anomalies to
be obvious, yet the
calibration problems may influence the derived values of $T$ and
$L$ in ways yet to be understood. 
We proceed anyway, with the caveat that all SSS spectra will
need to be revisited once the calibration is complete.  

With one exception,  blackbody models
provided acceptable fits; all values of $k\, T$ lie between
$50$ eV and $117$ eV, with 
$2.4 \times 10^{20}$ cm$^{-2} < N_H < 2.14 \times 10^{21}$ cm$^{-2}.$    
(See Table 4.) 
Thus, applied to real data, the conditions acted as the
simulations predict.  
The single source that required a 2-component fit was
M101-102, as observed in its high state during the long observation,  
yielding $> 8000$ photons. Even in this state,
only $\sim 2\%$ of the energy was received in photons with
energy $> 1.1$ keV, validating the selection procedure.
Three SSS-HR spectra are shown in Figure 2.  

It seems fair to assume that those SSS-HRs which provided
fewer counts are good SSS candidates. Indeed, the hardness ratios
of 
all of the SSS-HR sources are well within the limits set by Eqns.\, 1
through 4;
roughly $2/3$ of them have HR1$=$HR2$=-1.0.$

\subsection{Connections to Previous Work} 

 SSSs have been selected in $2$ of these
galaxies by other groups. Sarazin et al. (2001) identified
as SSSs all $3$ 
SSS-HRs.
Comparison with the work of Pence et al. (2001) on M101 is more difficult.
They selected $10$ SSSs, but only $7$ of them satisfy the HR conditions.
While it is not clear why $3$ of the SSSs identified by Pence et al.\, (2001)
do not satisfy the HR conditions, it is likely that the $5$ SSS-HR sources 
not selected by Pence et al.\, (2001) are slightly harder than the sources they
targeted. 
Indeed, the criteria used by Pence et al.\ (2001) required
SSSs to provide 
$100\%$ of their counts at energies below 0.8 keV, and
more than $2/3$ of their counts at energies below 0.5 keV (Pence, private
communication). The sources in Pence et al.\ (2001), should 
therefore preferentially have $k\, T$ near or below $50$ eV.
Indeed, the first three sources in Table 2 
are M101 sources that were
also identified as SSSs by Pence et al.\ (2001),
and all have estimated values of $k\, T$ below $75$ eV.
The fourth source    
in Table 2 is an M101  source that was not identified
as supersoft by Pence et al.\ 2001, presumably because it emits some
high-energy photons, in spite of the fact that
HR1 and HR2 are $-0.9$ and $-1.0,$ respectively.  

\begin{table*}
\caption{Spectral Fits to the Brightest SSSs}
{\centering
\footnotesize
\begin{tabular}{lccccccc}
\hline
\hline

ID& $N_H$& kT$^a$ & kT$_{RS}$$^b$ 
&$\chi^2_{\nu}$/dof& Flux\,$^c$ & $L_X$\,$^d$ & $L_{bol}$\,$^e$\\\\
  &  ($10^{21}$cm$^{-2}$)& (eV) & (keV) & & & &\\
\hline
M101-104 & $1.16^{+1.08}_{-0.44}$ & $67^{+8}_{-14}$ && 1.39/9
&2.11&0.74, 1.13 & 3.62, 5.57\\ 
M101-51 & $2.14^{+0.83}_{+0.19}$ & $50^{+3}_{-12}$ && 0.57/7
&6.19&2.16, 3.32 & 33.53, 51.62\\
M101-18 & 1 (fixed) & $71^{+18}_{-19}$ & &  0.96/13 & 1.02 & 0.35,
0.55&1.61, 2.47\\
M101-102 (1st)& $1.54^{+0.19}_{-0.31}$& $91^{+7}_{-4}$&
$0.74^{+0.04}_{-0.05}$& 1.18/75&82.67&28.8, 44.4& 67.94, 104.59\\
M101-102 (2nd) & $0.95^{+0.67}_{-0.55}$ & $76^{+11}_{-10}$ && 0.75/8&
20.7 & 7.22, 11.11& 35.0, 53.8\\ 
M83-50& $0.24^{+0.74}_{-0.24}$ & $66^{+13}_{-24}$ && 1.24/5 & 1.14 &
0.28, 0.30& 0.98, 1.03\\
M83-54& $0.95^{+1.38}_{-0.69}$ & $71^{+10}_{-16}$ && 0.66/7 & 2 &
0.50, 0.53&2.26, 2.39\\
M83-79 & $2.1^{+3.74}_{-1.53}$ & $86^{+27}_{-30}$ && 0.33/5 & 4 & 1,
1.1& 2.69, 2.85\\
M51-12 (1st) & $0.5^{+0.92}_{-0.46}$ & $102^{+12}_{-16}$ && 1.12/15 &
6.17&4.38, 5.20&11.86, 14.11\\
M51-12 (2nd) & $0.6^{+1.8}_{-0.6}$ & $117^{+26}_{-29}$ && 0.63/9
&4.82 &3.42, 4.07&9.13, 10.87\\
NGC4697 (3 sources)& $0.50^{+0.31}_{-0.37}$ & $81^{+11}_{-7}$ && 0.66/6
& 1.89&3.10, 12.28 &10.72, 42.51\\
\hline
\end{tabular}
\par

\medskip

\begin{minipage}{0.8\linewidth}
\footnotesize

NOTE.--- All quoted uncertainties are at the 90\% confidence
level. The ACISABS
model was applied to correct for the 
low-energy degradation of ACIS (see Kong et al. 2002).\\
$^a$ Blackbody temperature.\\
$^b$ Raymond-Smith temperature.\\
$^c$ Unabsorbed flux ($\times10^{-14}$ erg
cm$^{-2}$ s$^{-1}$) in 0.3--7 keV.\\
$^d$ Luminosity ($\times10^{38}$ erg s$^{-1}$) in 0.3--7
keV. The upper and lower limits quoted here are based on the
farthest and nearest distances, respectively,
shown in Table 2.\\
$^e$ Bolometric luminosity ($\times10^{38}$ erg s$^{-1}$).

\end{minipage}
\par
}
\end{table*}

\begin{figure*}[htb]
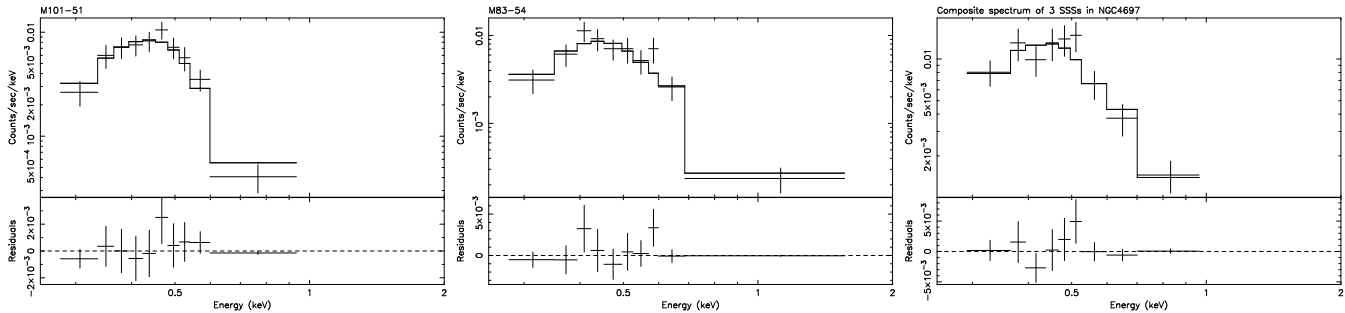

\rotatebox{-90}{\psfig{file=m101-51.ps,height=2.3in}}
\rotatebox{-90}{\psfig{file=m83-54.ps,height=2.3in}}
\rotatebox{-90}{\psfig{file=ngc4697sssspec.ps,height=2.3in}}
\caption{Blackbody models for M101-51 ($k\, T =50$ eV), M83-54 
($k\ T=71$ eV) and the composite spectrum of the 3 brightest SSSs in NGC4697
($k\ T=81$ eV).} 
\end{figure*}
\subsection{Variability}

If the SSSs we discover in other galaxies are X-ray binaries,
it is possible for them to exhibit variability on short time scales.
Although not every X-ray binary is highly variable, evidence of variability
on month-to-year time scales can immediately rule out the 
hypothesis that the source
of the SSS emission is a region as large as $\sim 0.1$ pc.
Note that variability within an observation can be observed for only
the brightest SSSs, such as M101-102. In most cases,
the detection 
of variability can best be accomplished by repeated observations
of the same galaxy.  

Three of the $4$ galaxies considered here have been observed
twice. M101 (M83) was observed for $\sim 100$ ksec ($\sim 50$ ksec)
 and then 
again for $\sim 10$ ksec ($\sim 10$ ksec). M51 was observed for $15$ ksec and
$26$ ksec. (See Table 2 for the dates of the observations)

\subsubsection{M51} 
Because the relative observing times are most similar for the 
$2$ M51 observations, we have created source lists for both
observations in
order to study the variability of the SSSs. 
The combination of the $2$ observations identified $4$ SSS-HR
sources, each of which was variable at some level.

One SSS-HR from the
$15$ ksec observation (M51-58-1) was not detected in the $26$ ksec 
observation; based on the detection of $21$ photons in
the short observation, we would predict $31$ photons in
the second observation; none were detected.  
One SSSs from the longer observation (M51-51-2) was not 
detected in the 
shorter observation; $10$ counts were predicted in the 
short observation.
M51-42-1 (14.7 counts; all in $S$) 
exhibits a clear change; it is detected (49.5 counts; M51-52-2)
and is soft in the second observation, but is not an SSS-HR. 
The brightest SSS in M51,
M51-12-1/M51-13-2, decreased in flux by $50\%$ between the observations,
and became slightly harder
(see Figure 3 and Table 2). 

The fact that some level of variability is exhibited by all of M51's SSS-HRs  
suggests that the majority of SSSs in M51 are not SNRs
or other extended emitters.

\subsubsection{M101}

In both M83 and M101, the shorter observation had a duration
only $0.1$ that of the longer observation; we can therefore check
for the variability of only the brightest sources.  
For the purposes of this paper we have checked the position of
each SSS-HR identified in the long observation of M101 
to determine if a soft source was detected during the $10$ ksec
observation.  Seven of the $20$ SSS-HR sources were  
detected and were soft during the $10$ ksec observation
(M101-18, 34, 51, 97, 101, 102, 104).  We have not determined
whether these $7$ sources satisfied the HR condition during the
short observation, but we do find that the observed count rates
during the short observation were consistent with those taken during the 
long observation in all but one case.
Among the $13$ long-observation 
sources that were not detected in the short observation,
the predicted numbers of  counts 
range from $3$ to $8$. In most cases
there were no photons within the PSF. 
It is very likely that some or even most of the 
non-detection indicate genuine variability.  
We can write the probability of the source having the same
luminosity as in the long observation but not being detected in the short
observation as $0.1\, p_i,$ with $p_i < 10;$ in most cases $1 < p_i < 5.$ 
The probability, $P,$ 
that all of thirteen sources actually had similar fluxes to those
observed in the first observation is therefore  $10^{-4} <P< 10^{-13}.$ 
Additional observations of long duration are required to 
quantify the level of variability among M101's SSS-HR sources. 

One of the most exceptional examples of SSS variability
found to date is provided by M101-102. In the long 
observation it is clearly ultraluminous (see also Mukai et al. 2002). In the short
observation the flux has declined by a factor of $\sim 4,$
due to a combination of changes in $L$ and $T.$ 
Furthermore, this source is clearly variable within each observation.

\subsubsection{M83}

The two brightest
SSS-HR sources (M83-54 and M83-79) were not detected in the
2001 short observation. Scaling for the exposure time, 
we would have expected 
12-15 counts
from each source. The lack of detection
 does not prove the sources  are transient, but
does indicate a significant decrease in flux.
We note that  
two non-SSS sources nearby each had a similar number of counts
in the long observation, and {\it are} 
detected again in the short exposure.
See Soria \& Wu (2003) for a more detailed
discussion of variability in M83.

\subsection{Spatial Distribution}

With our sample of $33$ SSSs in external galaxies, it is
becoming  possible to search for patterns in their spatial distribution
within the galaxies; such studies could shed light on
the nature and evolution of the sources.
Determinations of the relative numbers of SSSs in galactic
bulges may allow to us to test
the hypothesis that tidal disruption events should produce
SSSs that are the hot cores of stripped giants (Di~Stefano et al.\ 2001)   
There are, however, too few SSS-HRs in the central regions
of the $3$ spirals studied here to permit meaningful statistical 
tests at present.   Instead we focus on the disk regions, and
find an unexpected pattern. 

Superposing the positions of the SSSs on optical images of each of
the $3$ spiral galaxies (Figure $3$) produces an impression that a large
fraction of the SSSs are in the spiral arms.
In most cases, populations
found in the   spiral arms are young populations, with maximum ages on the
order of $10^8$ years.

The expected pattern was based on a clear expectation for
the nature of the SSSs themselves.
Specifically, a large portion of the SSS population
was expected to be composed of hot WDs.  
Since recent novae, hot central stars
of PNe and symbiotics have all been observed as SSSs,
and in fact comprise nearly half of all SSSs  with
optical IDs, we expected that
at least a fraction of the SSSs in other galaxies should be associated
with such systems. In novae and symbiotics, the donor stars
are not typically high-mass stars, and they should therefore not 
presently be
associated with young stellar populations. 
Furthermore, the most popular model   
for the remainder of the SSSs also invokes a binary with
a slightly evolved donor star with a mass between approximately
$1$ and $3\, M_\odot$.
SSSs described by these models should therefore also
be largely associated with populations of old or intermediate-age
stars, and should therefore be spread across the face of
spiral galaxies, and not found primarily near concentrations
of young stars.  

In Figure 4 we show the spatial
distribution of novae (red circles marked with
dates), PNe (yellow crosses) and SSS-HRs (marked red circles) 
in M101. 
As expected, the novae and PNe 
cover the disk of the galaxy.
For example, even though there are only 11 novae,
several are located away from the arms and also away
from regions with bright blue clumps.
The same is true of 
PNe. In contrast, the $16$ SSS-HRs are largely
absent from these open spaces, and tend to be located within $10''$
of bright blue clumps.    

At the distance of M101, $10''$ corresponds to roughly $200$ pc,
which is the distance a star traveling at $200$ km s$^{-1}$
can cross in $10^6$ yrs. If SSS-HRs are young systems,
we would therefore expect to find them within $\sim 10''$ of markers of
recent star formation, including bright stars, SNRs, HII regions, 
and OB associations.      
Therefore, more convincing than a visual impression of an association
between the locations of SSSs and the spiral arms,  would be
a meaningful statistical correlation between the
positions of the SSSs and the positions of markers of young
populations.
Work is underway to study these correlations in M101 and M83.
Below we mention some specific cases in which SSSs 
are close to optical markers in M101 and M83. The results of a
statistical analysis will be reported elsewhere.
In M101, one of the SSSs lies within $2''$ ($\sim 48$ pc) of
an SNR.  
Astrometric study is underway to 
assess the likelihood of a genuine identification.
Because the source has a relatively
low count rate, we do not have a spectrum. 

In addition, $4$
distinct SSS-HRs 
(M101-51,78,80,104),  
are within $10''$ of an HII region. 
We have spectra for $2$ of these sources: M101-104 and M101-51
are among the softest sources in our sample, with $k\, T$ of $67$ eV
and $50$ eV, respectively.

In M83 one of the SSS-HRs (M83-88) is
 located within $10''$ of an SNR (R. Soria 2002,
private communication). The impression that the SSS-HRs are
associated with
young objects is reinforced by examining their positions
relative to clumps in an H$_\alpha$ image of M83 (Figure $4$.)   

We note that our results are consistent with the only other map of
SSSs in a spiral galaxy that we are aware of (Swartz et al. 2002). 
While we cannot verify that the $9$ SSSs on their map satisfy the 
HR conditions, it is interesting to note that $6$ or $7$ of these
sources appear to be located in the spiral arms.  

If indeed we can establish a spatial correlation between SSSs 
and markers of young stellar populations, and if this
correlation is significantly larger than the correlations between 
PNe (and novae) and markers of young stellar populations,
then we will be forced to conclude that a substantial fraction
of SSSs are members of relatively young stellar populations.
The data so far clearly indicate that
the SSSs we are viewing in other galaxies are not primarily
members 
of old or intermediate-age  
populations.

\begin{figure*}[htb]
\centerline{
\psfig{file=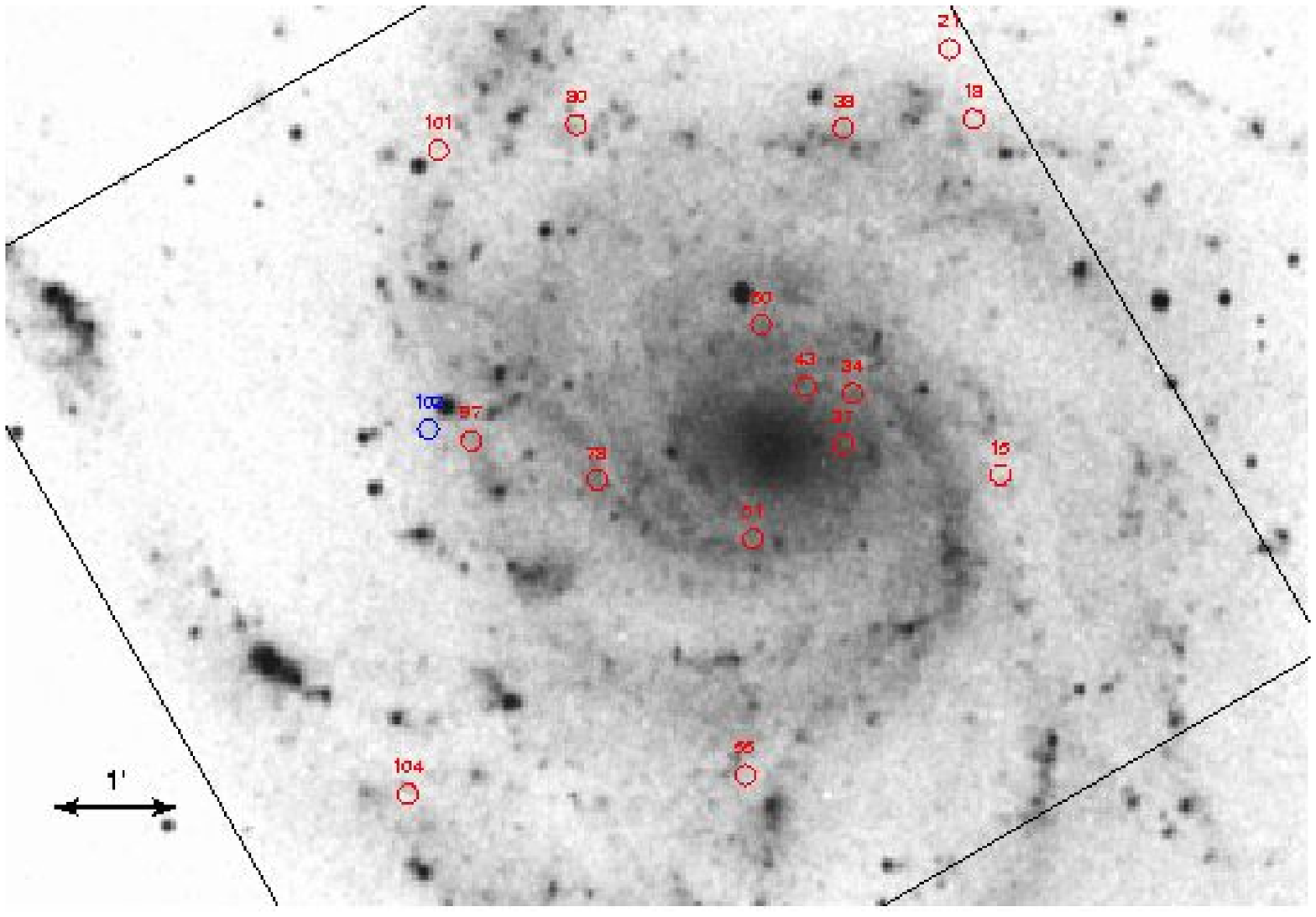,width=3.3in}
\hfill
\psfig{file=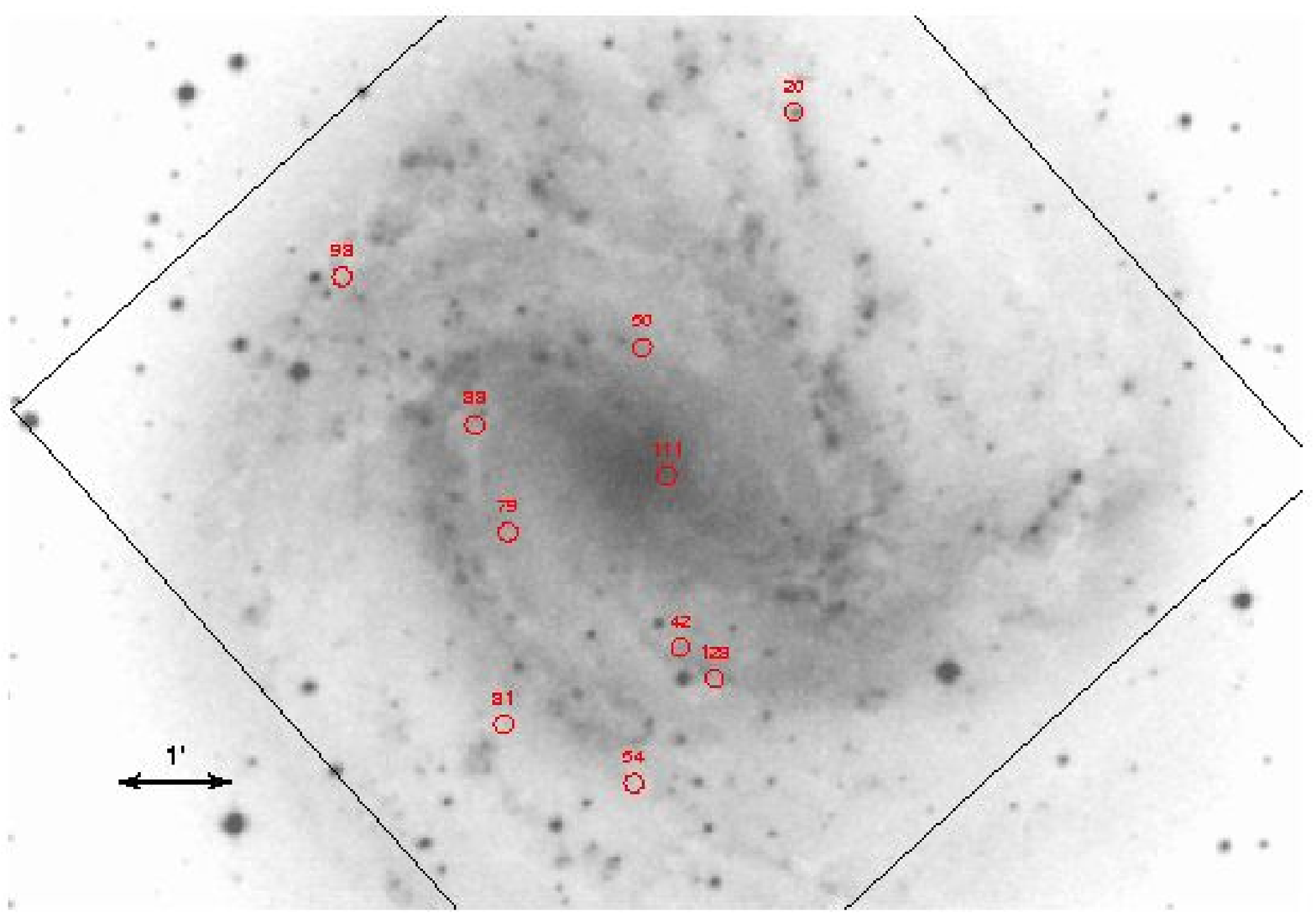,width=3.3in}
}
\centerline{
\psfig{file=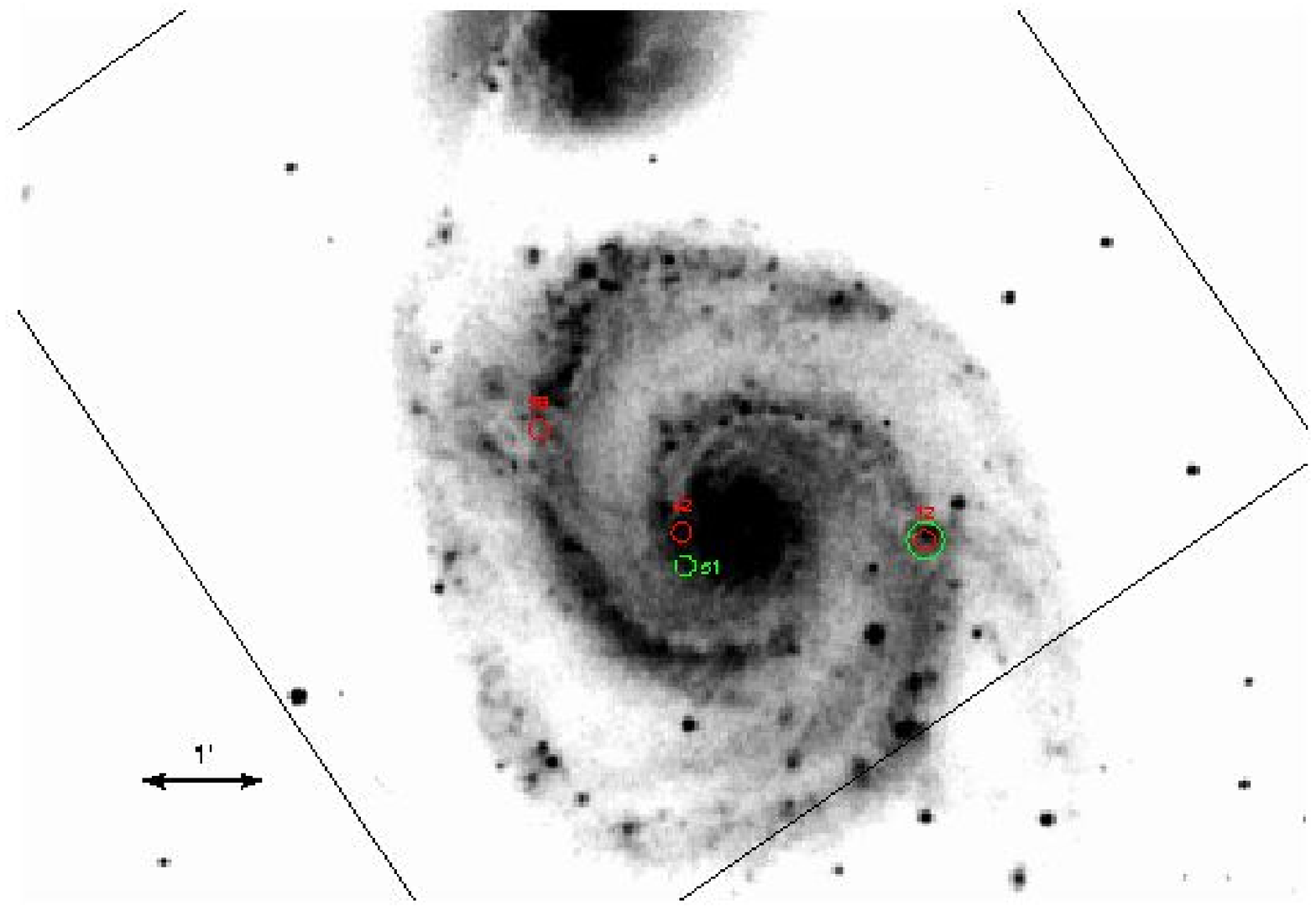,width=3.3in}
\hfill
\psfig{file=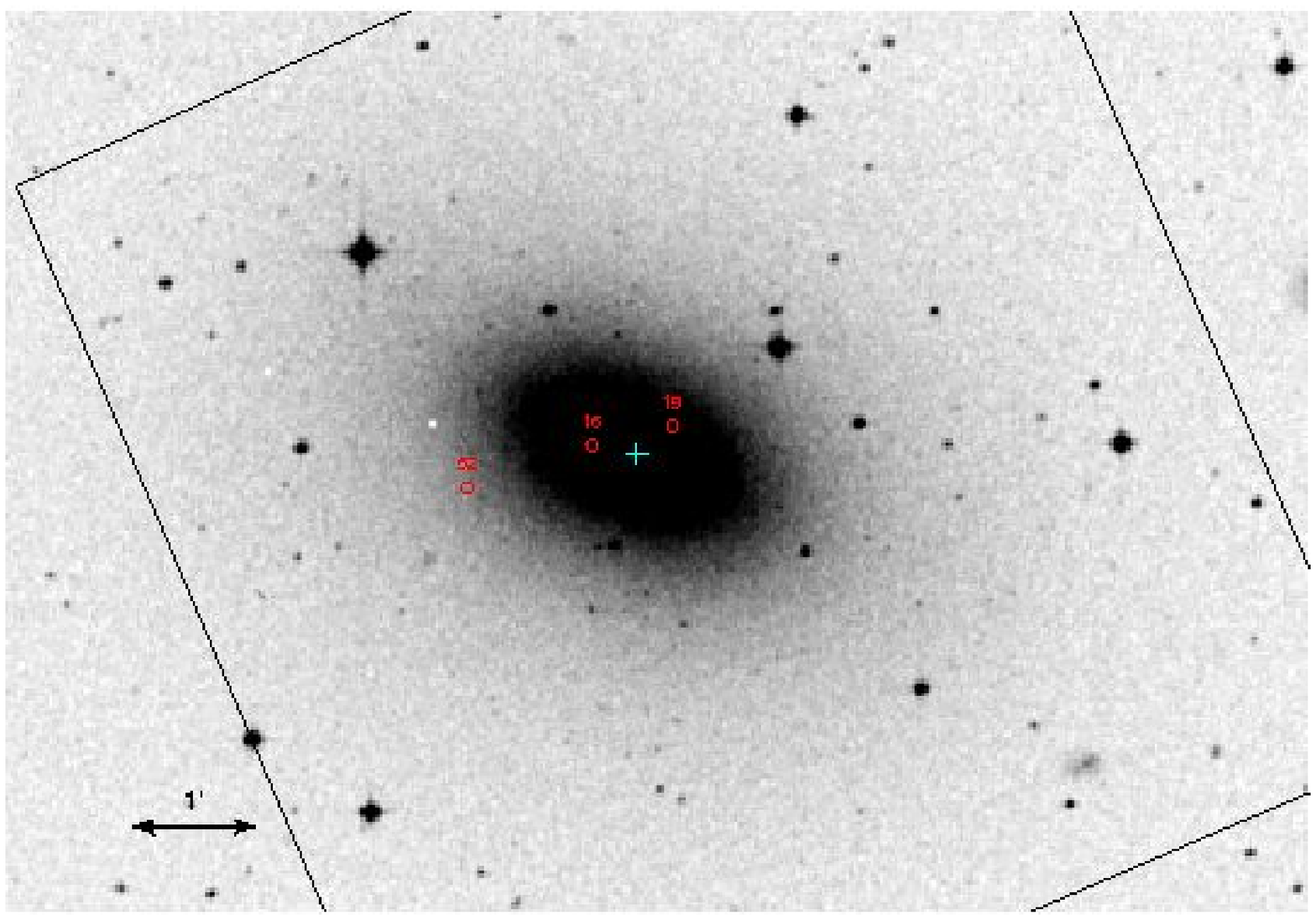,width=3.3in}
}
\caption{{\it Chandra} field-of-view (ACIS-S3; black line) overlaid on
the Digitized Sky Survey images of M101 (upper
left), 
M83 (upper right), M51 (lower left) and NGC4697 (lower right). Also
shown in the figures are the positions with source numbers (red
circle) of all
SSS-HR candidates. The ultraluminous SSS M101-102 is marked by blue
circle. In M51, those SSSs found in the second observation are marked
by green circle. Note that the central $\sim 10''$ of M51 is excluded. The
center of NGC4697 is marked by a blue cross. North is up, and east is to the left.}
\end{figure*} 

\begin{figure*}[htb]
\centerline{
\psfig{file=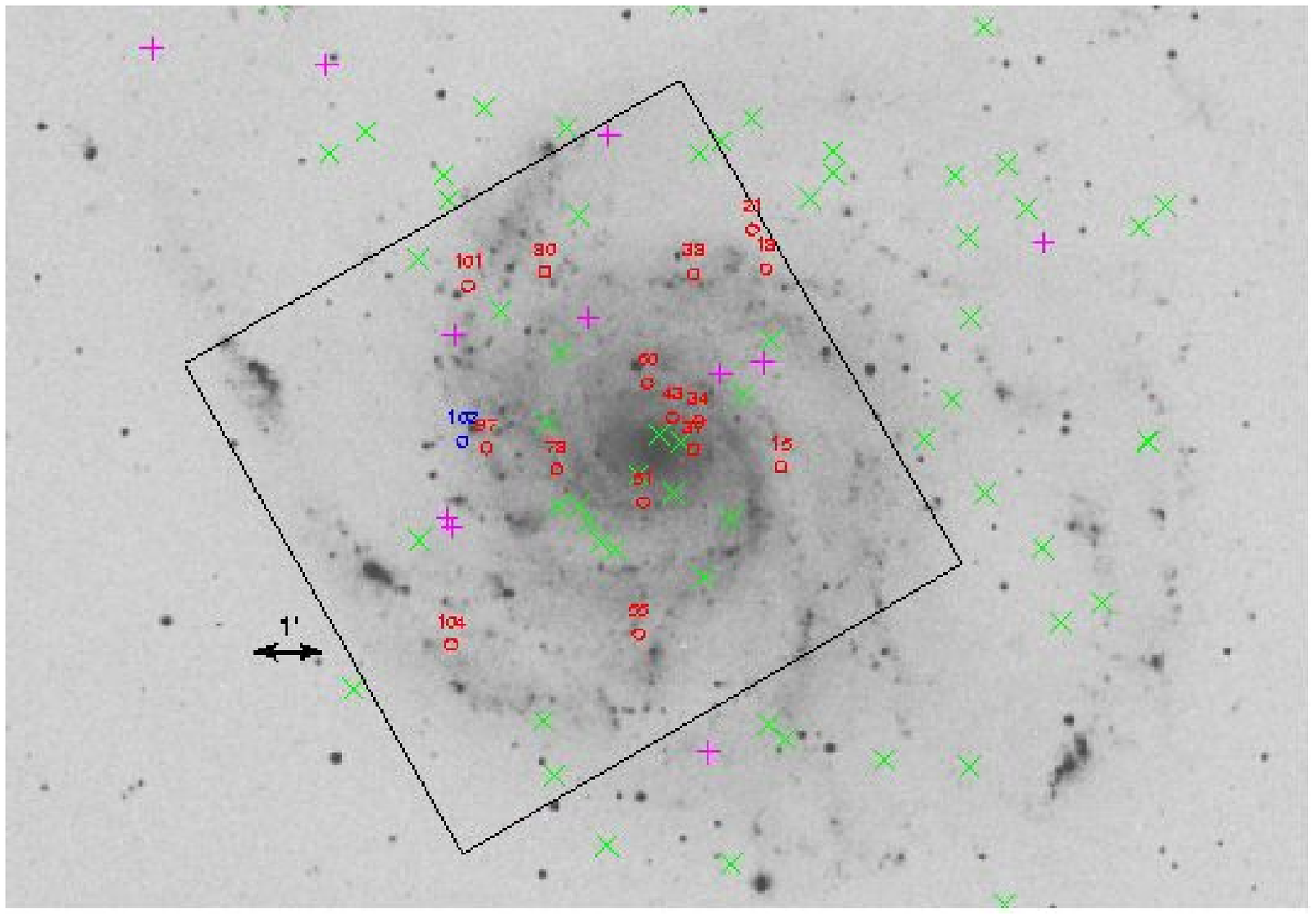,width=3.3in}
\hfill
\psfig{file=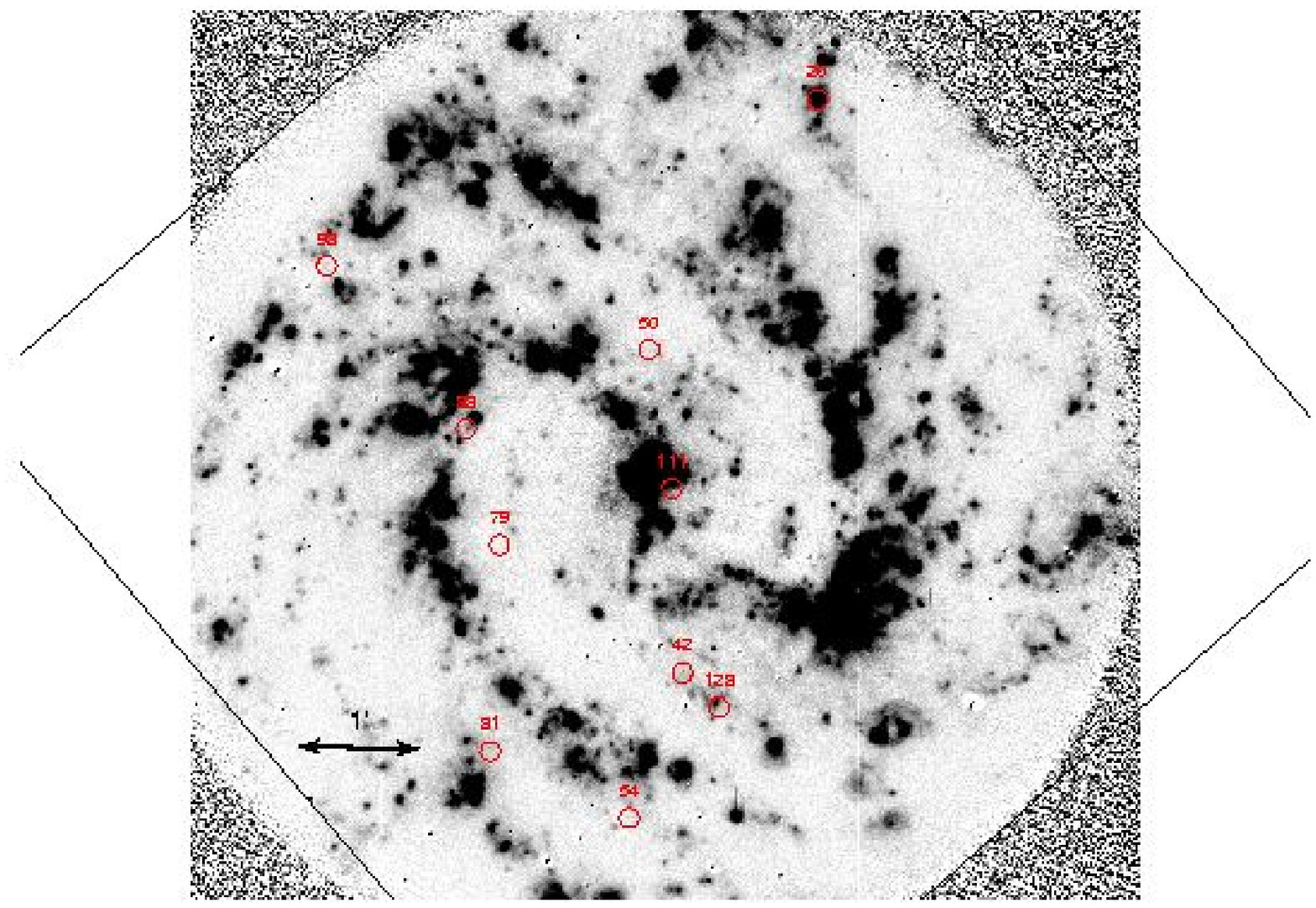,width=3.3in}
}
\caption{Left: The positions of PN (green crosses) and novae (purple crosses) overlaid on the Digitized Sky Survey image of M101. Symbols for SSSs are the sane as Figure 3. Right: H$\alpha$ image (provided by Stuart Ryder and Roberto Soria) of M83 with positions of SSSs overlaid (symbols same as Figure 3). North is up, and east is to the left.}
\end{figure*}

\begin{figure*}[htb]
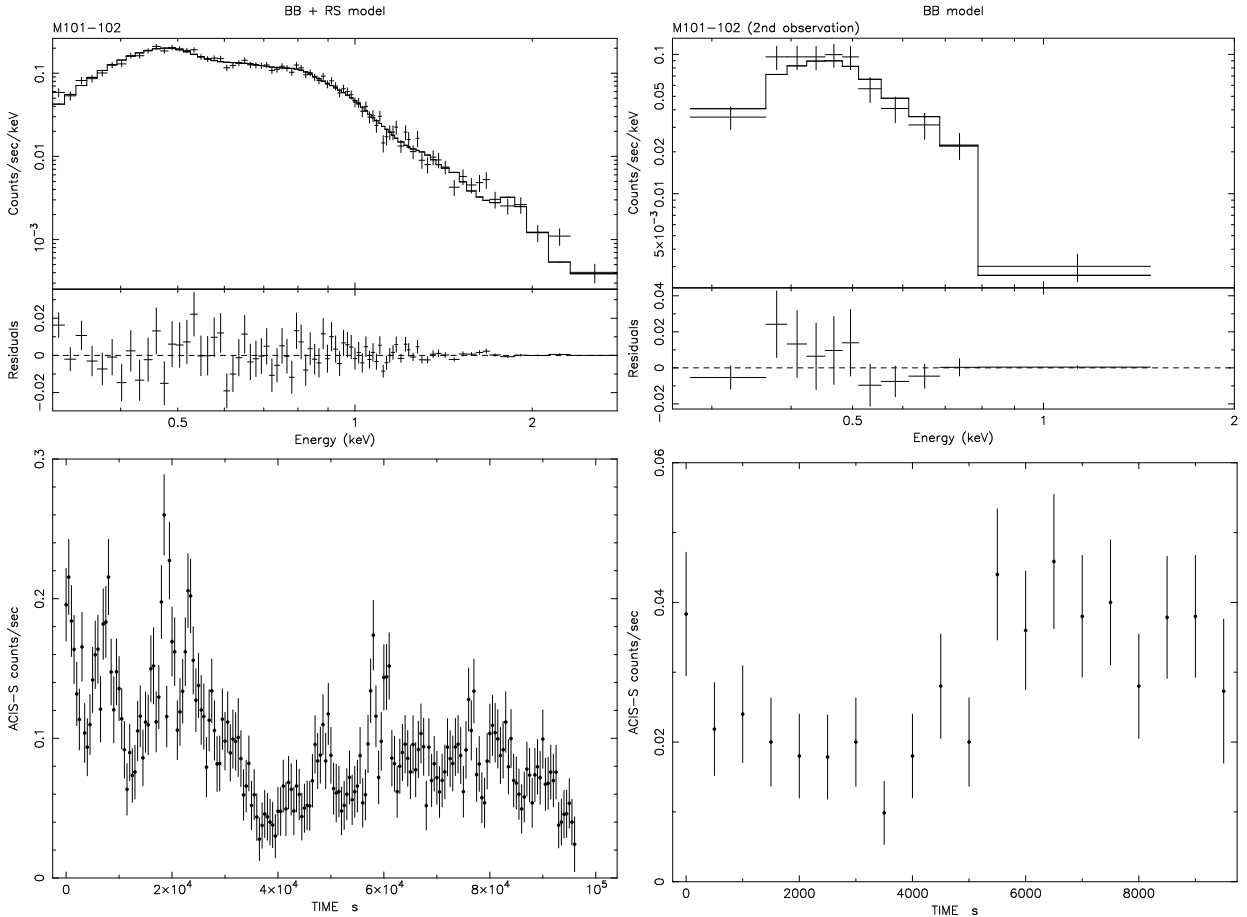

\centerline{
\rotatebox{-90}{\psfig{file=102.ps,height=3.2in}}
\rotatebox{-90}{\psfig{file=m101-102b.ps,height=3.2in}}
}
\centerline{
{\rotatebox{-90}{\psfig{file=m101-102a.lc.ps,height=3.2in}}}
{\rotatebox{-90}{\psfig{file=m101-102b.lc.ps,height=3.2in}}}
}
\caption{X-ray spectra for M101-102 during the first observation
(upper left; blackbody plus Raymond-Smith model)) and second observation
(upper right; 79 eV blackbody model). Lower panel is the
0.3-7 keV lightcurves (500 s time resolution) of the same sources. Note that the vertical axes have different scales.}
\end{figure*} 

\begin{figure*}[htb]
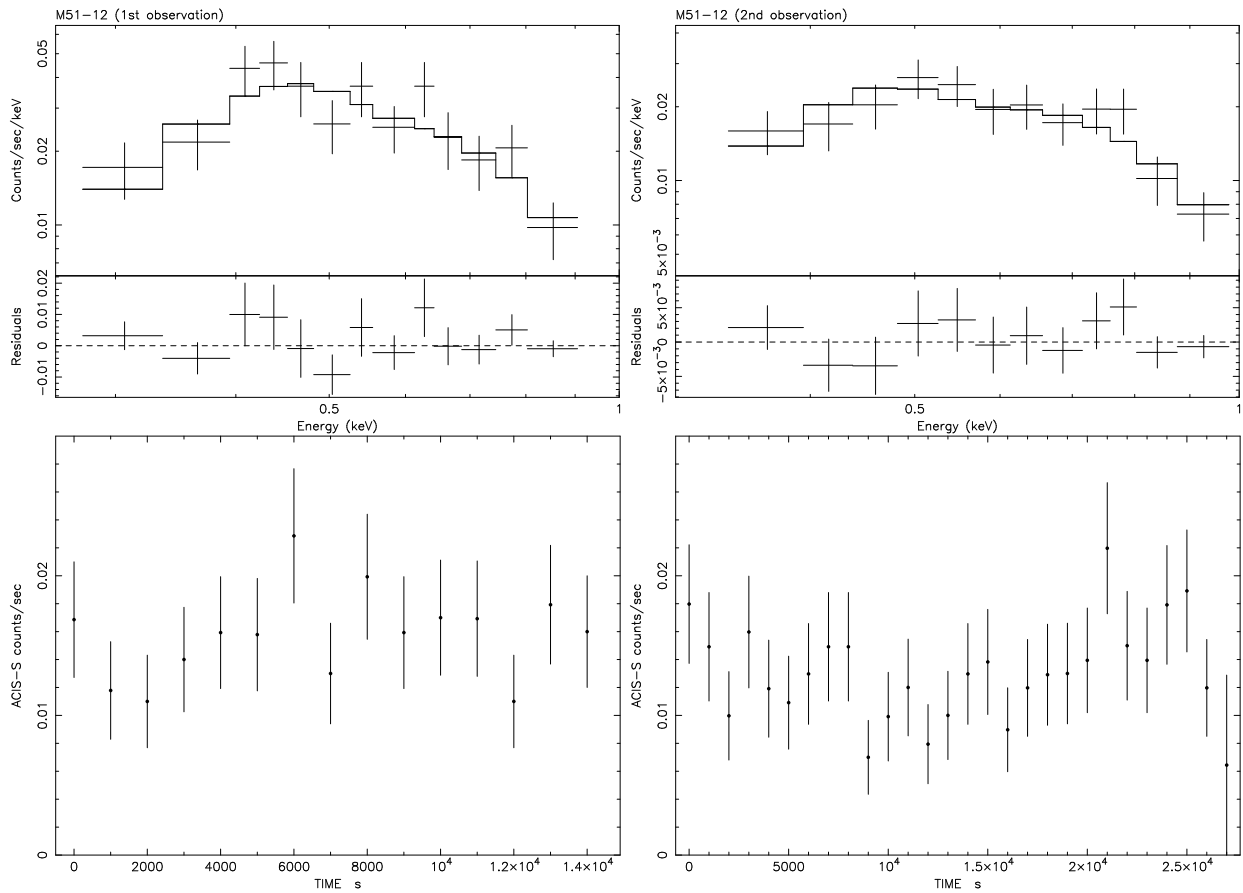

\centerline{
\rotatebox{-90}{\psfig{file=m51-12a.ps,height=3.2in}}
\rotatebox{-90}{\psfig{file=m51-12b.ps,height=3.2in}}
}
\centerline{
{\rotatebox{-90}{\psfig{file=m51-12a.lc.ps,height=3.2in}}}
{\rotatebox{-90}{\psfig{file=m51-12b.lc.ps,height=3.2in}}}
}
\caption{X-ray spectra for M51-12 during the first observation
(upper left; 102 eV blackbody model) and second observation
(upper right; 117 eV blackbody model). Lower panel is the
0.3-7 keV lightcurves (1000 s time resolution) of the same sources.}
\end{figure*}

\subsection{Luminosities}

In most X-ray studies of galaxies, luminosities are estimated for all
sources by using a standard spectral model,
generally a power law model,
to convert the measured count rate to a luminosity.
For SSSs, the count rate depends so strongly on the 
source temperature and on the size of the intervening gas column, 
that such a simple procedure does not produce meaningful results.
We have therefore made luminosity estimates only for the $11$ 
sources for which we have spectral fits (Table 4).

In general, meaningful spectral fits are possible for
just a small fraction  of all sources in 
galaxies more distant than $5$ Mpc. 
In NGC 4697, for example, $3\%$ ($10\%$) of all sources
provide more than roughly $300$ ($150$) counts;
in M101, $14\%$ ($24\%$) of all sources
provide more than roughly $300$ ($150$) counts. 
For SSSs, the effects of absorption can compound those of distance,
with low-T sources most affected.
A $40$ eV, $1 \times 10^{38}$ erg s$^{-1}$ source, 
located 8 Mpc away, and observed for 
$100$ ksec by {\it Chandra's} ACIS-S, will yield 
approximately $47,$ $8,$ and $0.1$  
photons if $N_H$ is $4 \times 10^{20}$ cm$^{-2},$
$1.6  \times 10^{21}$ cm$^{-2},$ and
$6.4 \times 10^{21}$ cm$^{-2},$ respectively.
The corresponding numbers of counts for a $55$ eV source
are $106,$ $23,$ and $0.4,$
respectively. Given these numbers, which were computed with the AO3 release of 
PIMMS,
, it is clear that
it will not be possible to obtain spectra for many extragalactic
SSSs with $k\, T < 55 eV.$ Even higher-temperature SSSs must be very luminous
if we are to collect a large number of photons: for an $80$ eV
$1\times 10^{38}$ erg s$^{-1}$ source located $8$ Mpc from us, 
the numbers of counts expected
for the $3$ values of $N_H$ given above are 
$227,$ $75,$ and $4,$ respectively. 

It is therefore remarkable that we were able to
extract spectra for $1/3$ of the SSSs found in the $4$
galaxies. 

\subsubsection{ULSs: Ultraluminous SSSs}

The most luminous  SSS in the galaxies we have studied is
M101-102. It's luminosity during the long exposure 
was clearly $> 10^{39}$ erg s$^{-1}$; thus, by most
definitions, it
qualified as an ultraluminous X-ray source. 
M101-102 is an ultraluminous SSS 
(ULS), and it is the only source in this data set
whose X-ray luminosity
alone is $> 10^{39}$ erg s$^{-1}.$  

Other sources with bolometric luminosities larger than $10^{39}$ ergs s$^{-1}$
are M101-51, M51-12 and, 
the $3$ SSS-HRs in NGC 4697 (if 
the true distance to that galaxy is close to the
maximum estimated distance).
These sources must, however, be viewed as weaker ULS
candidates, since (a) the number of photons received from
them was typically more than an order of magnitude
smaller than from M101-102, making the spectral fits less
certain, and (b) the derived value of the bolometric
luminosity is model-dependent.

Even if the estimates of the bolometric luminosities are too high
by an order of magnitude (unlikely to be the case for all of the
sources), M101-51, M51-12 and the SSS-HRs in
NGC 4697 
have luminosities near or above the 
Eddington limit for a $1.4\, M_\odot$ star.   

\subsubsection{Sources Near the Eddington Limit}

The estimated bolometric luminosity of $5$ additional sources
lies near the Eddington limit for a $1.4\, M_\odot$ star. 
If any
of these are NBWDs (whether or not the donor fills its Roche lobe),
they are particularly interesting, because they could be Type Ia supernova
progenitors with masses near the Chandrasekhar limit.

\subsubsection{Sources of Moderate Luminosity} 

Two-thirds of the SSS-HRs
provided only between $14$ and $80$ counts, with almost
half of them providing fewer than $30$ counts.
Scaling from the high-L systems, and taking the detection
limit for each galaxy into account, it is likely that typical
SSS-HRs  in our sample have luminosities in the range of a few times 
$10^{37}$ to $10^{38}$ erg s$^{-1}$. That is, 
unless their temperatures are considerably lower than $50$ eV,
or the intervening column is large,  
their luminosities are
similar to the luminosities of the local SSSs discovered
during the early years of ROSAT.\footnote{There
are a subset of local SSSs that are associated with CVs;
these, like the nova-like variable V751 Cyg (Greiner et al. 1999), are less luminous.
Such sources, recognized as SSSs only in recent years, are below our
detection threshold.}

\subsection{Total galactic populations} 

Based on a small number of ROSAT-identified SSSs,
galactic populations of SSSs
were estimated to be large, with $\sim 1000$ 
$L > 10^{37}$ ergs s$^{-1}$ sources in spiral galaxies such as
the Milky Way and M31 (\rd\ \& Rappaport 1994; Motch, Hasinger \& Pietsch 1994). 
In addition, populations numbering in the hundreds had been conjectured
to explain the diffuse soft X-ray emission in some elliptical galaxies
(Fabbiano, Kim, \&
 Trinchieri 1994).

Because of the effects of absorption, 
the SSS-HRs we have detected in each galaxy comprise a small fraction of all
of the SSSs within the galaxy. We detect primarily the hottest, brightest
SSSs.
This effect becomes even more pronounced for distant galaxies
and/or for (a) galaxies located along lines of sight
with large values of $N_H,$ (b) galaxies with large internal gas components,
or (c)  spiral galaxies with large inclination angles.

To estimate the size of the total population, we must determine
what fraction of all the sources we can (1) detect and (2) identify
as SSSs. The second requirement is more restrictive,
as typically more than $14$ photons are needed.
Assuming that at least $14$ photons are needed
we can compute, for each temperature
and each value of $N_H,$
the minimum luminosity of a detectable SSS in each galaxy.
We have done these  computations for each value of $T$ and $N_H$
used in the PIMMS simulations described in \S 2.

We have then seeded each of the $4$ galaxies with
the populations of SSSs 
used by \rd\ \& Rappaport (1994) to
estimate the fraction of SSSs that the {\sl ROSAT All-Sky Survey}
had discovered in the Galaxy, M31, and the Magellanic Clouds.
This tells us the fraction of all ``classical" SSSs
($k\, T < 100$ eV, $10^{37}$ erg s$^{-1} < L < 2 \times 10^{38}$ erg s$^{-1}$),
that can be detected and identified as SSSs in each galaxy. 
We have assumed that, if we receive $14$ photons
from a true SSS, there is a probability of $1/2$ that it will identify
the source as an SSS-HR, hence as an SSS. 

\noindent {\sl M101 and M83:\ } The fraction of sources we would
detect and identify as an SSS ranged from $1\%$ to $2.5\%$, with a median value of $1.5\%$.
This implies that each ``classical" SSS detected represents
between $40$ and $100$ M101 SSSs, with a median expectation value
of $\sim 66$. If we assume that roughly half of the 
SSS-HR sources in each of these two galaxies
represent ``classical" SSSs, then 
the number of ``classical" SSSs in each 
galaxy must lie between $200$ and $800.$    

\noindent {\sl M51 and NCG 4697:\ } 
Because of the greater distance to these two galaxies (particularly for 
NGC 4697), and because of the relatively high inclination of M51,
we can detect a much smaller fraction of all ``classical" SSSs
in these galaxies. We compute that the fraction of detectable SSSs
lies between $0.05\%$ and $0.12\%$ for M51 and between $0.03\%$ and $0.1\%$ for
NCG 4697. If, therefore, even one of the SSS-HRs in each
galaxy corresponds to a ``classical" 
SSS, the underlying population of SSSs in these  
galaxies is large, likely to be on the order of $1000.$
We note that, especially for these galaxies, it is difficult to 
judge whether the few SSS-HRs we have identified are similar to the
classical systems, or if they belong to a hotter and more luminous subclass
of SSSs.    
Even in the latter case, however, it is unlikely that the entire
population of SSSs happens to have luminosities and temperatures
high enough to bring them over the detection limit. 
It is therefore  
all but certain
that the SSS population consists of a significantly larger number
of sources than we have been able to detect.

In all cases, a large portion of the population of SSSs,
the portion corresponding to low luminosities 
($< \sim 10^{37}$ erg s$^{-1}$) and/or low-temperatures ($k\, T < \sim 30$ eV),
is unconstrained. It is therefore worth noting that, in our galaxy,
some CV systems have been identified as low-L SSSs; the numbers of
such systems in our own Galaxy may be ten times larger than the
numbers of ``classical" SSSs.  In galaxies beyond the Local Group,
we can only hope to detect such sources indirectly, e.g., by 
observations of diffuse soft emission.

In order to improve the estimates for that portion of
the SSS population (the 
hotter and more luminous 
sources) that {\it can} be directly detected and/or constrained 
through {\it Chandra} and {\it XMM} observations,
several steps, including the following,
 can be taken: 
(1) more sensitive
sampling of the values of $N_H$ in the simulations,
(2) measurements of $N_H$ in the galaxies, tying the location of
the detected sources to the local values of $N_H$.  

At present it seems likely that each of the galaxies we have studied
houses a population of SSSs that consists of at least a few hundred sources
with $L > 10^{38}$ erg s$^{-1}$. .
Additional observations and more sophisticated analyses 
will improve this estimate.  

\section{Interpretation }

\subsection{The HR Conditions}

Supersoft X-ray sources form a large and intriguing component
of galactic X-ray source populations. As {\it Chandra} and {\it XMM}
discover more SSSs in external galaxies, comparative studies will
provide important clues to the natures and physical characteristics
of the sources. Such studies require a uniform selection procedure.
In this paper we take a first step toward developing a systematic
procedure to select SSSs from a galactic population of X-ray sources.

In the absence of a single well-tested physical model, we start with
an empirical spectral definition of SSSs. Our definition (\S 1.1) includes
all X-ray sources emitting $90\%$ of their flux in photons with
energies $< 1.1$ keV.

To identify SSSs in external galaxies, 
selection criteria are needed to assess sources providing too
few counts for meaningful spectral fits--i.e., the vast majority of
all sources. In this paper we    
have introduced a strict set of criteria, the ``HR conditions'',
which preferentially select SSSs
with $k\ T < 100$ eV.
Indeed, 
SSS-HR's, i.e., sources satisfying the HR conditions,
should have spectral characteristics most similar to
those of the SSSs discovered locally (in the Magellanic Clouds and Milky Way),
with limited contributions from sources with $k\, T > 100$ eV
and virtually no contamination 
from sources with $k\, T > 175$ eV.

The clear advantage of the HR conditions is that
they are conservative. We therefore expect that applying them will provide insight into the 
nature of sources with spectra 
like those of the flagship sources discovered so far in
the Magellanic Clouds and Milky Way. 
In some ways, the 
study of SSS-HRs in $4$ galaxies has verified expectations based on
the local sample of SSSs, but this study has also produced
some surprises.

\subsection{Results from $4$ galaxies}

Applying these conservative 
conditions to {\it Chandra} data from $4$ galaxies 
allows us to establish that SSSs are an important
component of galactic X-ray binary populations.
Based on observations of $3-16$ SSS-HRs in each of the $4$
galaxies we studied, we estimate that 
the minimum number of SSSs with $L_x>10^{37}$ erg s$^{-1}$
in typical galaxies is likely to be in the hundreds.
The numbers of lower-$L$ sources remain  
unconstrained by the observations carried out to date.
In addition we appear to have discovered a sub-population of 
more luminous SSSs that    
are not present in the Magellanic Clouds and Milky Way.

\subsubsection{The most luminous SSSs} 
 
The spiral galaxies in our sample, and
perhaps the elliptical as well,
appear to house small populations of SSSs that are
ultraluminous. These ULSs could very well be 
accreting intermediate-mass BHs. Their luminosities and temperatures
are consistent with what is predicted for
accreting BHs with masses between roughly
$100\, M_\odot$ and $1000\, M_\odot$.  

Others of the sources appear not to be ultraluminous, but
do have luminosities that are near-Eddington or super-Eddington 
for a $1.4\, M_\odot$ object.
Some of these could be accreting WDs with masses approaching the Chandrasekhar mass (Di\,Stefano 2003, in preparation). 

\subsubsection{Young SSSs?}
 
The spatial distribution of SSS-HRs is not consistent with
a population dominated by old stellar systems.
At present it seems likely that, at least in spiral galaxies,
the  population of SSSs includes some systems with
ages $< 10^8$ years. While some SNRs may exhibit SSS-like emission,
the SSS-HRs which we have been able to test for variability   
appear not to be SNRs.
At least a significant fraction are therefore likely to 
be X-ray binaries.

The age of a binary when it emits X-rays is generally determined
by the age of the donor star when it either fills its Roche lobe
or emits a large enough stellar wind that its compact companion
can receive mass at  
rates in the range $10^{-9}-10^{-7} M_\odot$ yr$^{-1}$. A Roche-lobe filling donor can transfer mass at these rates to a WD in a dynamically stable way only if it is not too much more massive than the WD. Masses of $\sim 1-3 M_{\odot}$ are allowed. Such systems should not be primarily associated with the spiral arms.  
In order for such rates to be achieved when the
mass transfer occurs through a stellar wind, the donor  must be
very evolved. 
Only stars with masses greater than $\sim 5\, M_\odot$
can become so evolved in times shorter than $10^8$ years.
If the accretor is a WD, the system would be a symbiotic;
symbiotics typically have donors of lower mass (see Kenyon 1986). 
On the other hand, high-mass X-ray binaries (HMXBs)
in which the donor
is a high-mass star and the accretor is a neutron star or BH are
well known. Indeed, the X-ray pulsar with the luminous soft
component, RX J0059.2-7138, has been conjectured to be an HMXB 
(Hughes 1994).   
It is very likely, therefore, that at least some SSSs  
found near star-forming regions have BH  or neutron star accretors.

\subsection{Status of the WD Models} 

Although the spatial distribution of SSSs in the galaxies we have
studied so far appears to be indicative of a younger population
than we expected to find, it would be premature to conclude that the WD models
of SSSs are not correct. It seems more likely that we have discovered new 
populations that supplement those we had hoped to find.

For example, within our own Galaxy, we would be unable to discover soft sources,
even luminous soft sources, located in star forming regions.
This study of M101, M83, and M51, which are
all located behind relatively small gas columns, and
which are less inclined than M31, has provided the first opportunity to
systematically search for SSSs among young stellar populations.  

Even if our conjecture is correct, i.e., some of the SSSs we have discovered are
not hot WDs, some SSSs in these 4 galaxies could be hot WDs.
Further
observations and even work with existing data may be able to
determine the relative contributions of SSSs in young and old stellar
systems. (See \S 5.2.)

\section{Prospects}

\subsection{Selection Criteria}

If the fact that the HR conditions are conservative selectors of SSSs
is their strength, it is also a weakness that prevents
them from identifying a significant fraction of all SSSs.
In tests based on simulations, the HR conditions 
fail to select about $1/2$  of all SSSs.
(See Table 1 and Figure 1.) The fraction of missed SSSs could be larger,
and in fact is almost certainly progressively larger in {\it Chandra}
 observations taken 
progressively later than the time of launch. The number of photons
in the S band will, by 2004, be a factor of $5$ smaller for some sources 
than predicted in the simulations we have used. 
(See the recent AO5 PIMMS release: for a hydrogen column of
$4\times 10^{20}$ cm$^{-2}$, the S counts predicted for AO5 are
$16\%$, $21\%$, and $30\%$ of what they were in the AO3 predictions
for a $40$ eV, $55$ eV, and $85$ eV source,
respectively.)   
  Clearly a more comprehensive
set of criteria are needed. We will return to this issue in a
separate paper (\rd\ \& Kong 2003).

\subsection{The Physical Nature of SSSs} 
This first comparative look at SSSs in external galaxies
raises intriguing questions about the nature of the sources.  
The following observations will help 
determine the answers.

\noindent {\bf (1)} Repeated X-ray observations.
These will allow us to (a) test the hypothesis that some of the SSSs
are SNRs, (b) identify SSSs which are recent novae, (c) determine
the duty cycles of SSSs which are X-ray binaries, and to (d) discover
SSSs of lower luminosity by merging multiple images. 
We note that {\it Chandra} observations should be conducted
as soon as possible to avoid the worst effects of the degradation in low-energy sensitivity.
 
\noindent {\bf (2)} Repeated observations at optical and UV wavelengths.
These could identify novae and any other events associated with SSS
behavior. They also provide the data necessary to study the stellar
populations with which SSSs are associated.  

\noindent {\bf (3)} Correlation studies.
It is difficult to establish optical IDs at distances of several Mpc,
since $1''$ corresponds to roughly $4\, D$ pc, where $D$ is the
distance to the galaxy, expressed in units of Mpc.
Nevertheless, we can hope to establish the types of stellar populations 
that tend to be associated with SSSs. Preliminary studies of M101 and M83, for example,
seem to indicate that the positions of SSSs are more 
strongly 
correlated with markers of young stellar populations than are the 
positions of recent novae, but that they are less correlated with
some markers of young stellar populations than the markers
among themselves (e.g., the correlation between SNR and HII regions
may be stronger than the correlation between SSSs and HII regions).
While these effects have yet to be quantified, they might suggest
that some SSSs are members of young populations, while others are,
as expected, 
members of older stellar populations.      
Furthermore, if SSSs tend to be associated with
objects that have special optical properties (e.g., if SSSs are
bright in UV), then it might be possible to establish some IDs.
If, for example, $5$ SSSs in one galaxy are associated with UV bright regions,
the probability that this is a chance occurrence can be assessed.  
If it is small, then it is likely that some of the associations are
real. We can then compare with IDs that have been established for
local SSSs, to determine if the types of objects with which SSSs 
in distant galaxies are associated are similar to those associated
with local and better studied SSSs.  

The combination of these steps taken over an interval of years
will help us to understand the physical nature of the
 various sub-populations that comprise the
class of luminous supersoft X-ray sources. 

\begin{acknowledgements}
We are grateful to 
Jochen Greiner, Roy Kilgard, Miriam Krauss, Koji Mukai, William Pence,  
Paul Plucinsky, Andrea Prestwich, Douglas Swartz, Roberto Soria, Harvey Tananbaum,
Kinwah Wu, and Andreas Zezas 
for stimulating discussions and comments. We thank
Roberto Soria for providing pre-publication
source lists for M83.   
We are grateful to Paul Green and the ChaMP collaboration
for providing
blank field source data. 
This research has made use of the
electronic catalog of supersoft X-ray 
sources available at URL http://www.aip.de/$\sim$jcg/sss/ssscat.html
and maintained by J. Greiner.    
RD would like to thank the Aspen
Center for Physics for providing a stimulating environment,
and the participants of the workshop on  
{\it Compact Object Populations in External Galaxies}  
 for insightful comments.  
This work was supported by NASA under AR1-2005B, GO1-2022X,
and an LTSA grant, NAG5-10705.  
\end{acknowledgements}

\end{document}